\documentclass[trackchanges]{aastex7}

\begin{document}

\title{Planetary Nebulae in the eROSITA eRASS1 catalog}

\author[0009-0005-6854-9619]{Haoyang Yuan}
\affiliation{Laboratory for Space Research, The University of Hong Kong, Hong Kong}
\affiliation{Department of Physics, The University of Hong Kong, Hong Kong}
\email{yuanaa@connect.hku.hk}

\author[0000-0000-0000-0003,sname=Europe,gname=Martin]{Martin A.\ Guerrero}
\affiliation{Instituto de Astrof\'\i sica de Andaluc\'\i a, IAA-CSIC, Glorieta de la Astronom\'\i a S/N, Granada E-1800, Spain}
\email{mar@iaa.es}

\author[0000-0002-2062-0173]{Quentin Parker}
\affiliation{Laboratory for Space Research, The University of Hong Kong, Hong Kong}
\email[show]{quentinp@hku.hk}

\author[0000-0002-6752-2909]{Rodolfo Montez Jr.}
\affiliation{Center for Astrophysics $\vert$\ Harvard\ \&\ Smithsonian, Cambridge, MA, USA}
\email{rodolfo.montez@cfa.harvard.edu}

\collaboration{all}{The HKU-IAA PN collaboration}

\begin{abstract}

Some planetary nebulae (PNe) host X-ray-emitting hot bubbles shaped by stellar wind interactions 
and/or harbor X-ray-emitting central stars due to accretion, shocks within their fast stellar winds, 
or even chromospheric emission from binary companions.  In both cases, the properties of the X-ray emission 
critically probe late-stages of stellar evolution for such low- and intermediate-mass stars. 
While extant Chandra and XMM-Newton observations have detected  X-ray emission in PNe, the numbers 
known remain very small ($\sim40$) compared to the overall Galactic PNe population ($\sim4000$). 
We have initiated a project aimed at increasing the sample of known PNe with X-ray emission using 
both current and new space-based X-ray telescopes such as the Einstein probe. 
To further investigate their X-ray properties to elucidate what drives current X-ray PN detections, 
we have cross-searched the SRG {\it eROSITA-DE} eRASS1 source catalogue and Hong Kong (HASH) PNe Database. 
Five known X-ray PNe have been detected (Abell\,30, NGC\,2392, NGC\,3242, NGC\,5315, and LoTr\,5), 
two new X-ray PNe are revealed (IC\,1297 and NGC\,2867), one (K\,1-27) is removed from previous X-ray 
compilations, and another 11 previously detected X-ray emitting PNe are not recovered. 
A comparison of the X-ray flux of detected and undetected X-ray PNe reveals that eROSITA eRASS1 is 
sensitive to PNe with X-ray fluxes larger than $\approx2\times10^{-14}$ erg~cm$^{-2}$~s$^{-1}$.  
The frequency of occurrence is $\simeq$0.5\% among the 1430 HASH True PNe in the eRASS1 footprint.

\end{abstract}


\section{A short history of Planetary Nebulae with X-Ray emission} 

Planetary nebulae (PNe) are a short-lived phase in the late evolutionary stages of low- and intermediate-mass stars.  
Since the first detection of X-ray emission from NGC\,1360 by EXOSAT \citep{deKorte1985}, more than 40 years ago, 
PNe have been selectively targeted by any available X-ray mission \citep[e.g., ROSAT,][]
{Guerrero+2000}. This has steadily but only gradually increased their number until the ambitious 
\emph{Chandra Planetary Nebulae Survey} (\emph{ChanPlaNS}) large programs targeted a large fraction of 
all PNe closer than 1.5 kpc. This resulted in a major increase in the number 
of X-ray PNe up to $\simeq$40 \citep{ChanPlaNS1, Freeman+2014, Montez+2015}. This has led to unprecedented 
progress in the understanding of the nature of X-ray sources in PNe.  

PNe can exhibit point and extended sources of X-rays. The former are associated with their central stars (CSPNe).  
Soft X-ray emission $\leq0.3$ keV, originates from photospheric emission from hot stellar photospheres. 
However, there are many cases when their X-ray emission is harder, $\geq$0.5 keV.  In those cases, the X-ray emission 
can be attributed to shocks in winds, as in OB stars \citep{LucyWhite1980}, but also to 
companions, either to the chromospheric emission from a late-type companion 
\citep[as in the case of LoTr\,5,][]{Montez+2010} or to accretion onto the 
CSPN \citep[as can be the case of NGC\,2392,][]{Guerrero2019eskimo}.  

The observed diffuse X-ray emission can actually be expected in the framework of the interacting stellar winds 
model (ISW) of PN formation \citep{Kwok+1978}.  The interaction of the dense and slow stellar wind ejected 
during the asymptotic giant branch (AGB) phase with the strong fast stellar 
wind from the CSPN produces a reverse shock heating the stellar wind to X-ray-emitting temperatures 
$\approx10^7$~K as in wind-blown bubbles \citep{Weaver+1977}. Heat-conduction, photo-evaporation, and turbulent 
mixing are expected to increase the density and emissivity of the X-ray emitting material while lowering its temperature 
down to several $10^6$~K \citep{TA2016}. The diffuse X-ray emission from these hot bubbles is thus soft and enclosed 
in most cases by sharp inner rims of some young, compact and dense PNe. 

Chandra and XMM-Newton have produced, for the first time, decent X-ray images and spectra of PNe. 
However, targeted observations have only been obtained for a very limited sample out of 
the 3957 known True (2787), Likely (459), and Possible (711) Galactic PNe, as recorded in the HASH PNe database \citep{HASH}.  
Moreover, their X-ray observations have been biased towards high surface brightness and well-studied PNe to 
increase the chances of detection and their value to broad, multi-wavelength comparisons of these often well 
studied PNe.  X-ray all-sky surveys, on the other hand, can probe the whole sample of Galactic PNe to assess 
the effects of distance, extinction, binarity, extant physical processes, and evolutionary state on the X-ray 
detectability and X-ray properties of all PNe. Previously this was only available for the low-resolution ROSAT 
All-Sky Survey (RASS), with a lower flux limit of a few times $10^{-13}$ erg~cm$^{-2}$~s$^{-1}$ \citep{RASS} 
compared to more recent X-ray telescopes like Chandra and XMM-Newton. ROSAT data are both limited and too shallow 
to allow this investigation. However, the partial release of the {\it eROSITA} all-sky survey 
\citep[the {\it eROSITA-DE} eRASS1,][]{eRosita}, with an increase in sensitivity of at least a factor of 10 
over the ROSAT all-sky survey, provides a unique opportunity to search for new X-ray-emitting PNe and to derive 
global properties for their X-ray emission. 

We present here a search for new X-ray PNe using the {\it eROSITA-DE} eRASS1 source catalog.
The cross-correlation between the known population of Galactic PNe and the sky footprint covered 
by {\it eROSITA-DE} eRASS1 is described in \S2. The results of this search is presented in \S3 and the 
implications for overall X-ray properties of PNe are discussed in \S4.  
The conclusions and expectations for new X-ray missions, as the Einstein Probe, are laid out in \S5.

\section{The sample of Planetary Nebulae and the eRosita eRASS1 catalog}
\subsection{The HASH Database: The comparison sample of Galactic PNe}


For comparison purposes with the X-ray data, this study uses the best available PNe optical images and spectroscopy from the Hong 
Kong/Australian Astronomical Observatory/Strasbourg Observatory H-alpha Planetary Nebula research platform database 
\citep[HASH, see][]{HASH}. This is further supplemented by new, very deep, narrow-band imaging from Peter Goodhew \footnote{
https://www.imagingdeepspace.com/peter-goodhew-planetary-nebulae-images.html
}, one of our amateur collaborators that are now regularly 
providing the best such PNe images currently available \citep[e.g.,][]{2022A&A...666A.152L}. 
HASH contains the consolidated discoveries and data for all known Galactic PNe in 
terms of multi-wavelength imagery, spectroscopy and other key parameters (angular size, morphology, central star 
identifications, and more) from compiling over 200 years of cataloguing and PNe discovery work in the field. 
 PNe central coordinates are derived from a fit to the overall geometric shape of the PNe, independent of any 
putative CSPN. For the vast majority of PNe, this is straightforward. Such PNe geometric centroid positions are usually 
very close to the position of well-identified CSPNe. In a few cases, there is a modest (few arcseconds) offset, 
sometimes seen in asymmetrically shaped PNe that may be interacting with the ISM.
For further details see \citet{2006MNRAS.373...79P},  \citet{2016JPhCS.728c2008P} and \citet{2022FrASS...9.5287P}.

\subsection{The eRosita and eRASS1 X-ray source catalogue of the Western Galactic Hemisphere ($l \gtrsim 180^o$)}

{\it eROSITA} (the extended Roentgen Survey with an Imaging Telescope Array) is an array of seven Wolter-I 
X-ray telescope modules, each equipped with 54 mirror shells and with an outer diameter of 36~cm on board 
the Russian-German Spektrum Roentgen Gamma (SRG) spacecraft. 
With a field of view of 1.03~degrees, it provides a wide field survey capability with 
high-throughput X-ray spectroscopic imaging in the 0.2-8 keV energy range with energy resolution $\simeq$80 eV at 1.5 keV 
\citep{eRosita}. Launched on 2019 July 13, its main goal is conducting a comprehensive X-ray survey of the entire sky, concluding its 
initial 184-day all-sky survey on 2020 June 11. Although primarily intended to map dark matter and Galaxy clusters, it has proven useful 
for any astrophysical source that emits sufficient X-ray flux to fall above the survey sensitivity limits. eRosita has a nominal all-sky flux 
sensitivity limit of 
$5\times10^{-14}$ erg~cm$^{-2}$~s$^{-1}$that improves at high ecliptic latitudes due to the all-sky survey 
scanning pattern. 
Unlike Chandra and XMM-Newton, intended as X-ray observatories targeting specific sources, eRosita provides not just 
Galactic plane coverage but all sky survey capabilities and so greater source discovery opportunities at the relevant sensitivities available.

The first data release of the Western\footnote{
Data rights are split by Galactic longitude ($l$) and latitude ($b$).  
Data with $-0.05576432^\circ < l <179.94423568^\circ$ (Eastern Galactic hemisphere) belong to the Russian 
consortium, while data with $179.94423568^\circ < l < 359.94423568^\circ$ (Western Galactic hemisphere) belong 
to the German consortium and are the only ones currently available and provided by eROSITA-DE.} 
Galactic hemisphere of the {\it eROSITA} all-sky survey (the {\it eROSITA-DE} eRASS1, 
hereafter eRASS1) was made public on 2024 January 31 \citep[][]{Merloni2024}.  eRASS1 provides calibrated data products and 
source catalogs processed using the {\it eROSITA} standard pipeline. Here we will use the eRASS1 Main catalog, with calibrated 
event files and count numbers, count rates, flux estimates, detection likelihood, observing time, and quality flags for 903521 point 
sources and 26682 extended sources detected in the 0.2-2.3 keV band. The survey is also sensitive in the harder X-ray range of 2.3-5 keV.

\subsection{Cross-correlation of the Planetary Nebulae sample with the eRASS1 catalog}
We cross correlated the positional information for all true (T), likely (L) and possibly (P) Galactic PNe in HASH with sources in the 
eRASS1 Main catalog.  Considering the typical soft X-ray emission exhibited by PNe, the main catalogue used is that for the soft 
0.2-2.3 keV {\it eROSITA} ML1 band. Given the typical eRASS1 imaging resolution (HEW$\approx30^{\prime\prime}$) and the possible crowding 
at the location of some PNe in the Galactic Plane, the search was initially limited to X-ray sources with a positional coincidence within 
$20^{\prime\prime}$ of a HASH Galactic PN. This resulted in the 10 eRASS1 PN candidates listed in Tables~\ref{tbl:pn_lst_ok}, 
\ref{tbl:pn_lst_new}, and \ref{tbl:pn_lst_pos}. 
The results are presented below. If we relax this conservative limit to $45^{\prime\prime}$ a further 7 potential matches to known Galactic PNe were uncovered. 
On examination none of these are plausible matches. Six of the PNe were compact with no overlap and clear separation between the outermost contours derived from the low X-ray source counts and the PNe. One PN Abell~26 is well resolved with an angular size of 38~arsec. Here, the outermost X-ray contours impinge on the western edge of the PNe while the centre of the PNe and X-ray source are offset by $\sim30$-arseconds. This is a poor match compared to all others plausible matches we have presented and is discounted here.

\section{Results}
\begin{figure}[t!]
\centering

\includegraphics[width=0.25\textwidth]{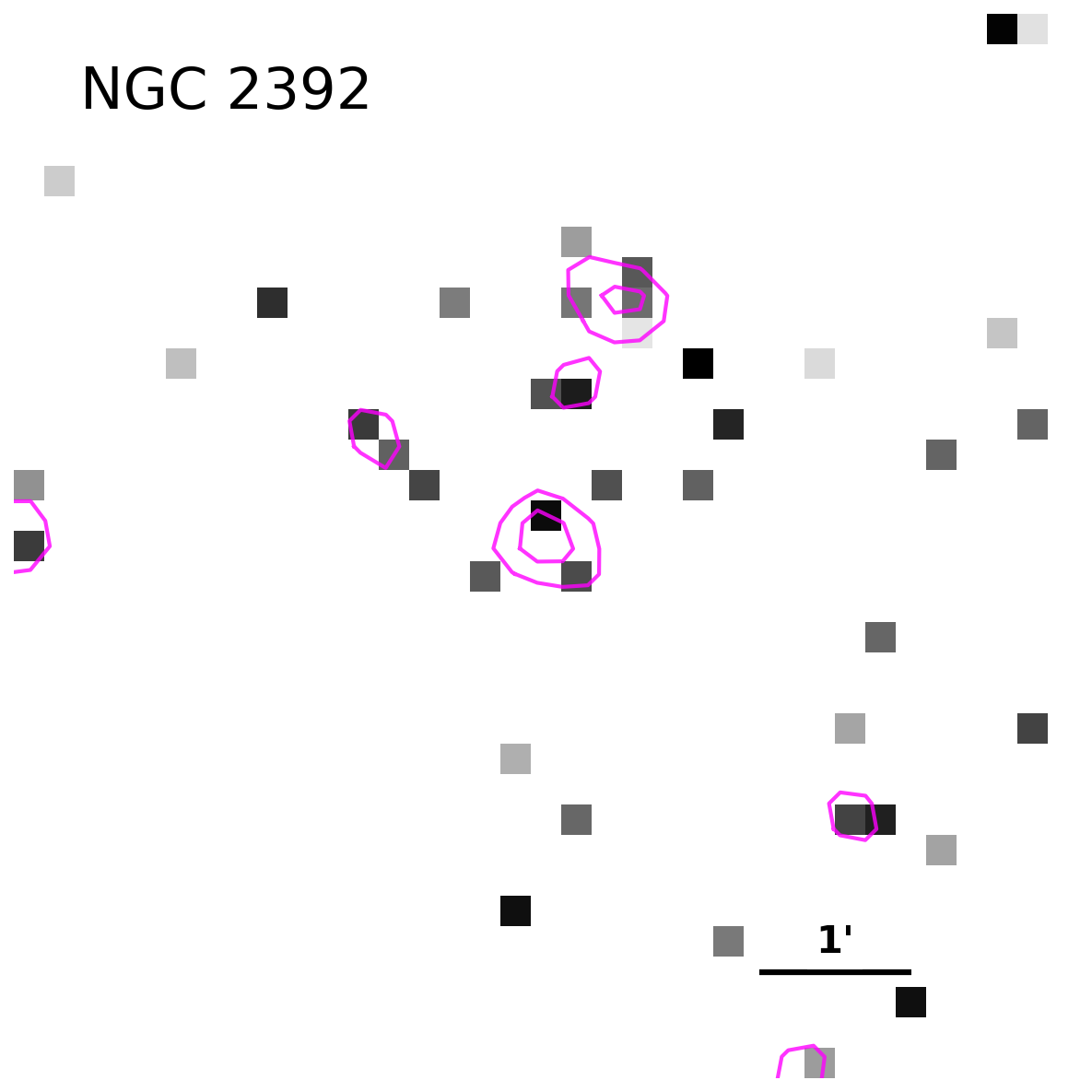}
\includegraphics[width=0.25\textwidth]{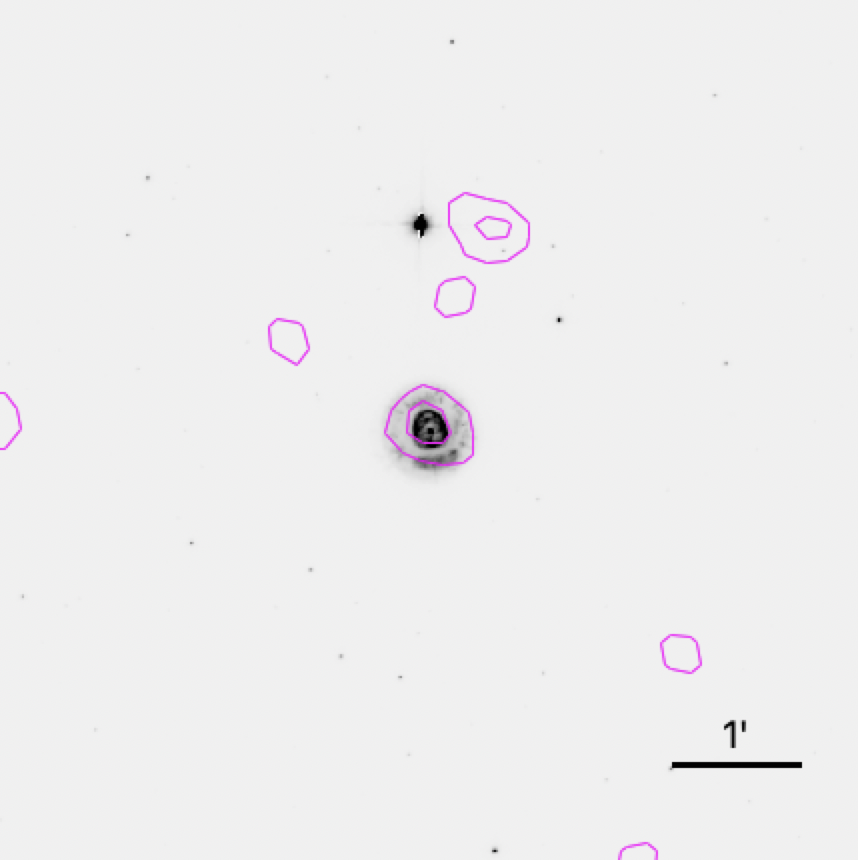}
\includegraphics[width=0.25\textwidth]{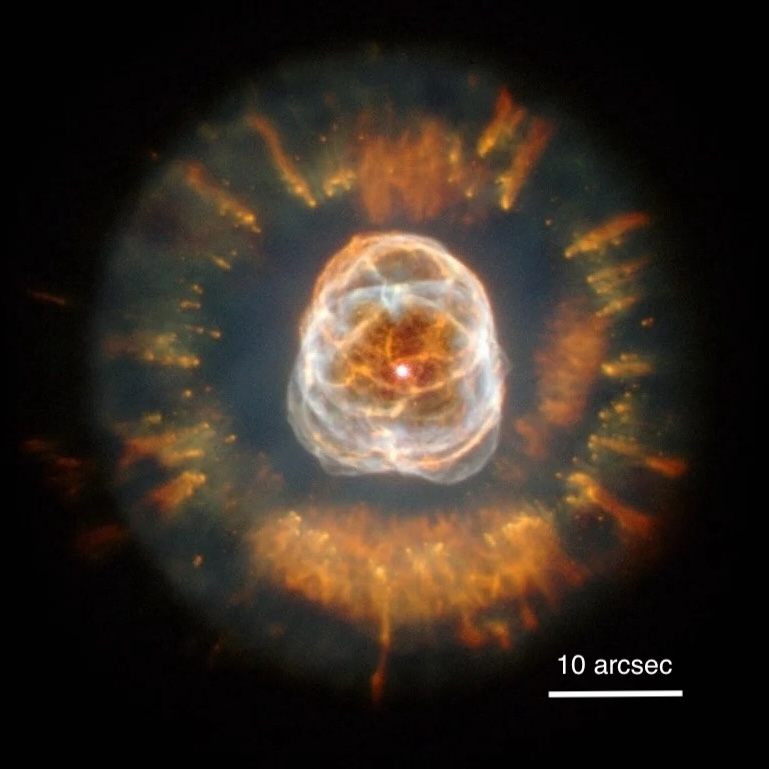} \\
\includegraphics[width=0.25\textwidth]{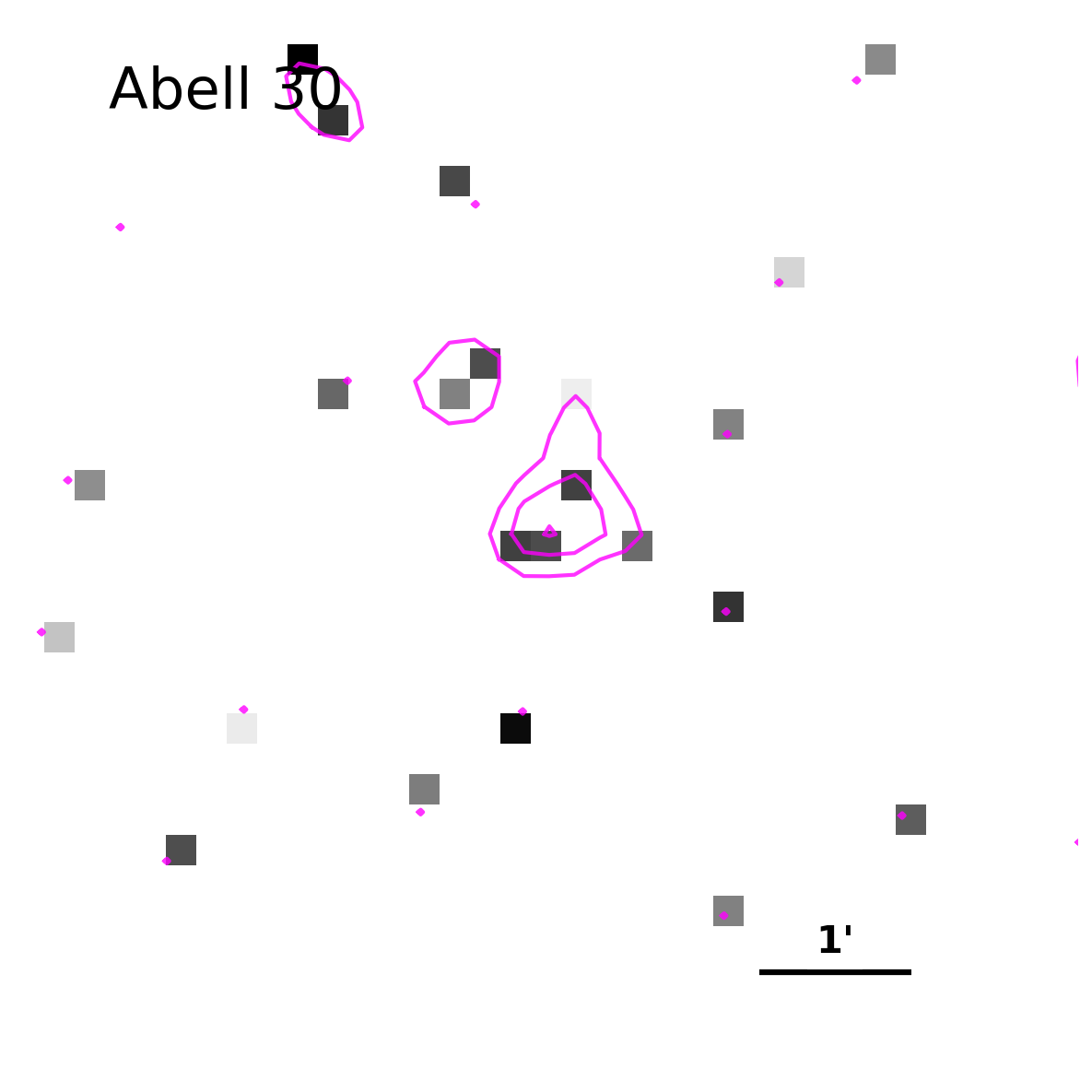}
\includegraphics[width=0.25\textwidth]{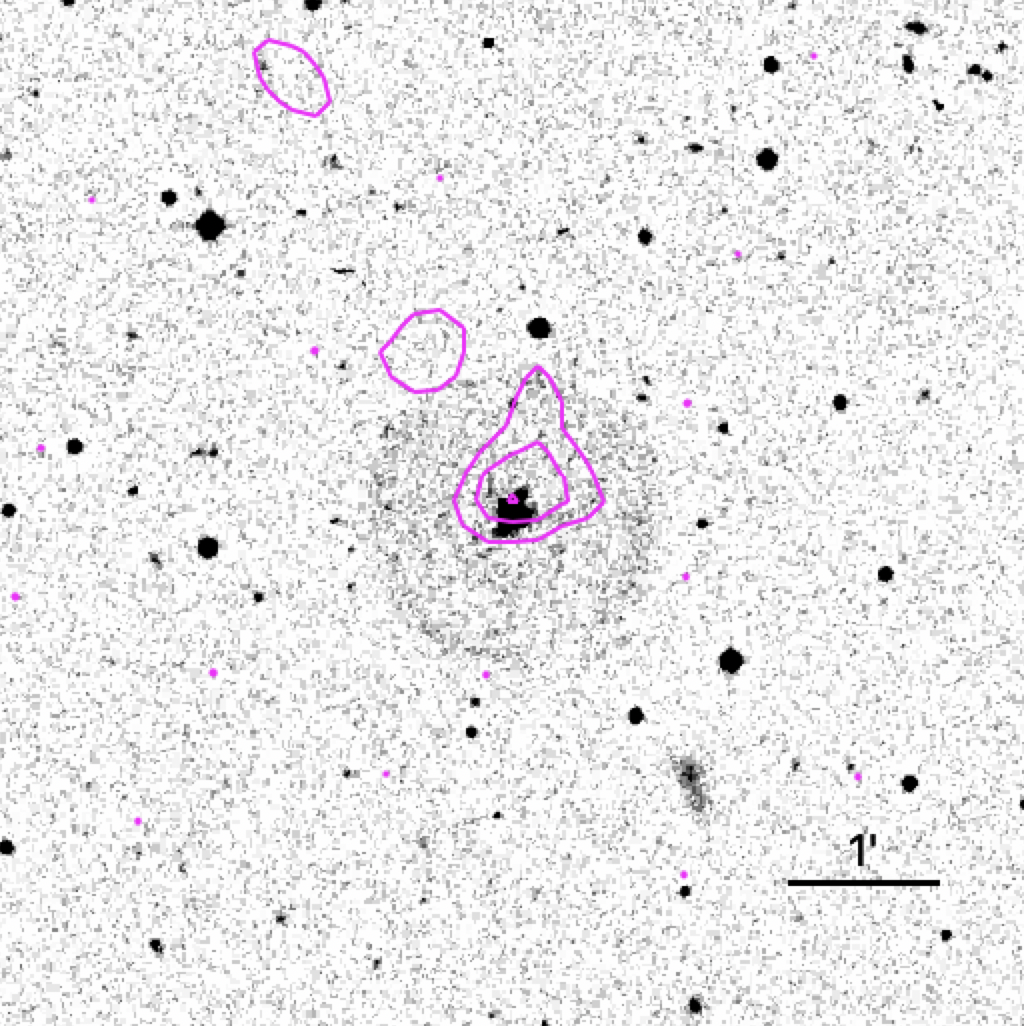}
\includegraphics[width=0.25\textwidth]{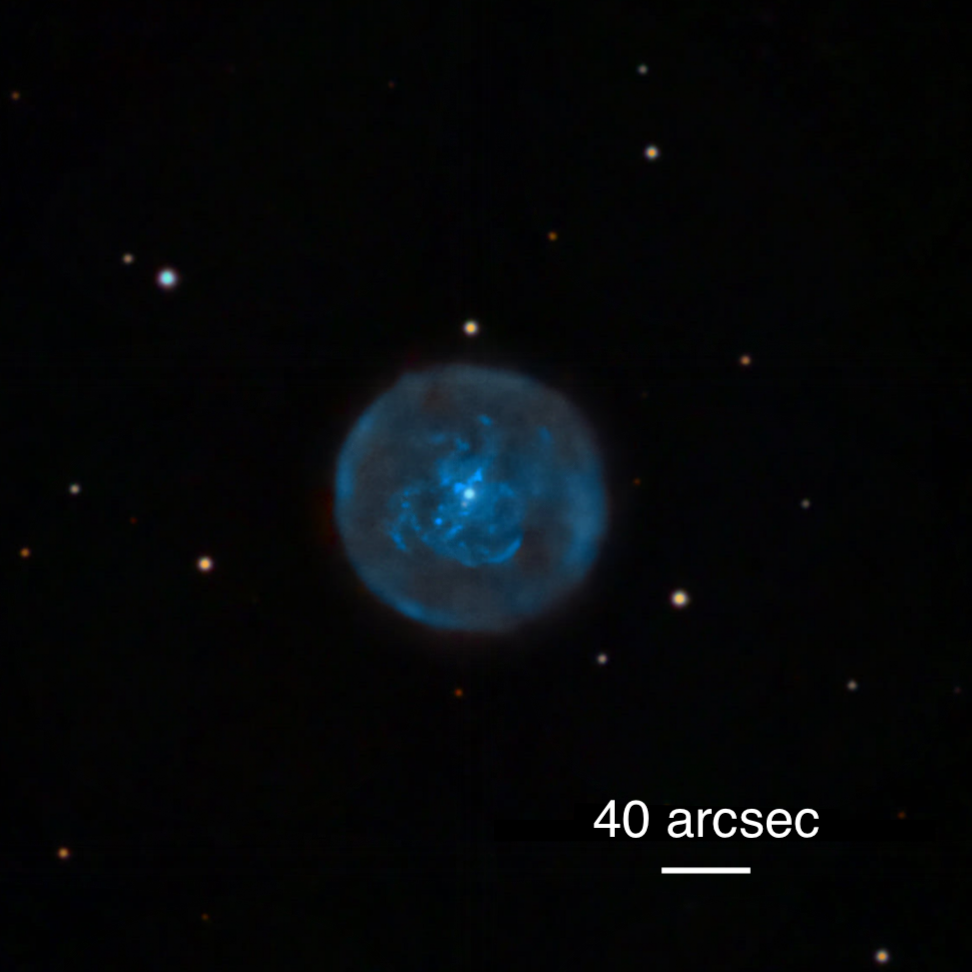} 
\includegraphics[width=0.25\textwidth]{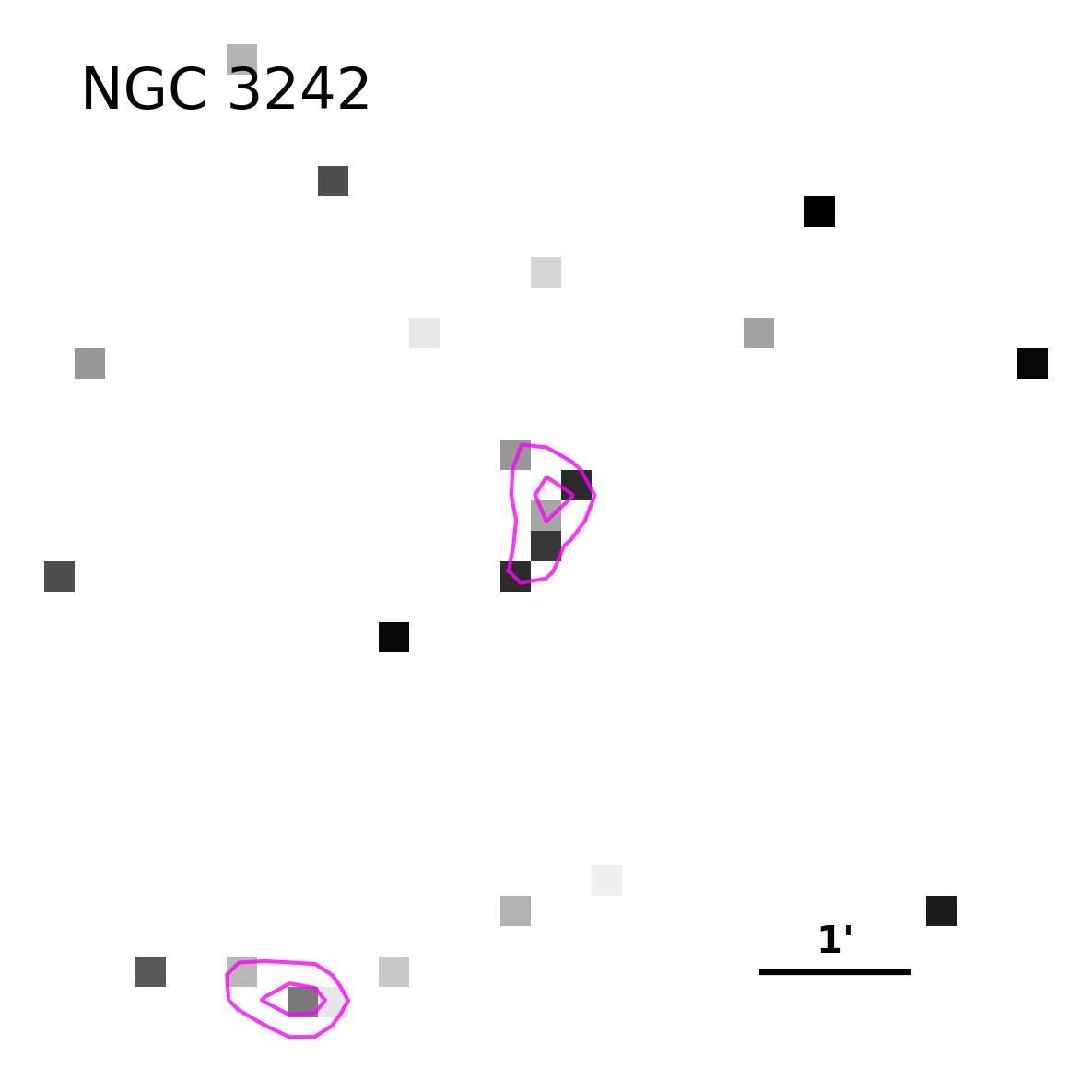}
\includegraphics[width=0.25\textwidth]{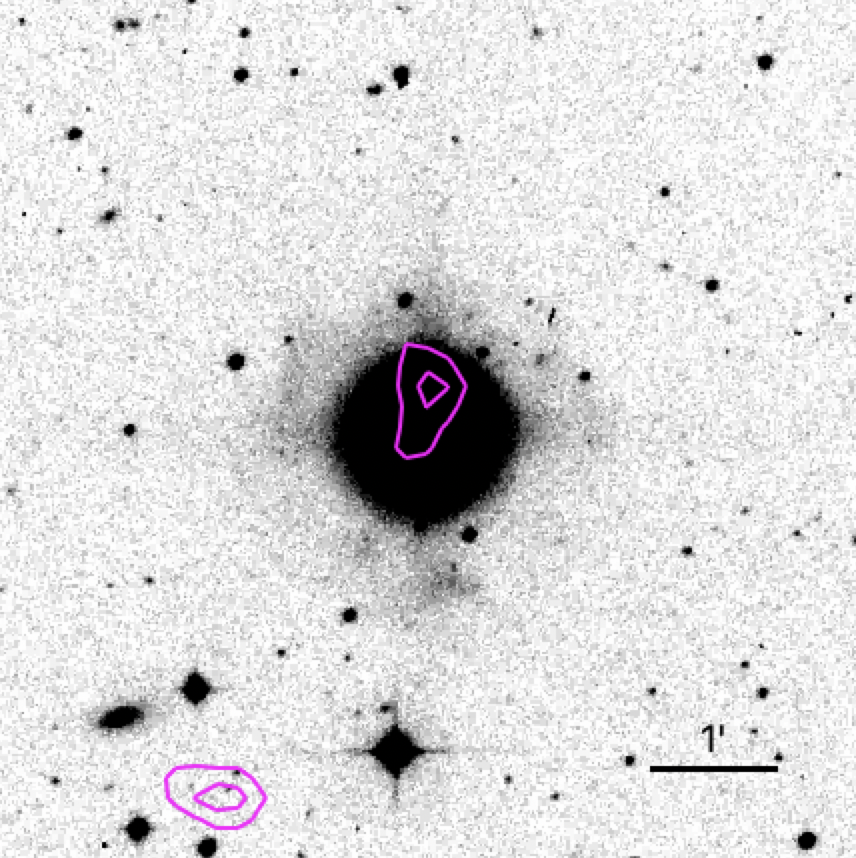}
\includegraphics[width=0.25\textwidth]{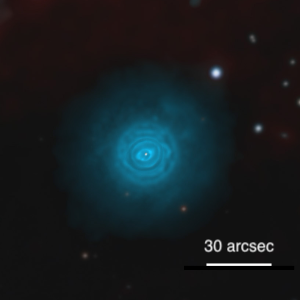} 
\includegraphics[width=0.25\textwidth]{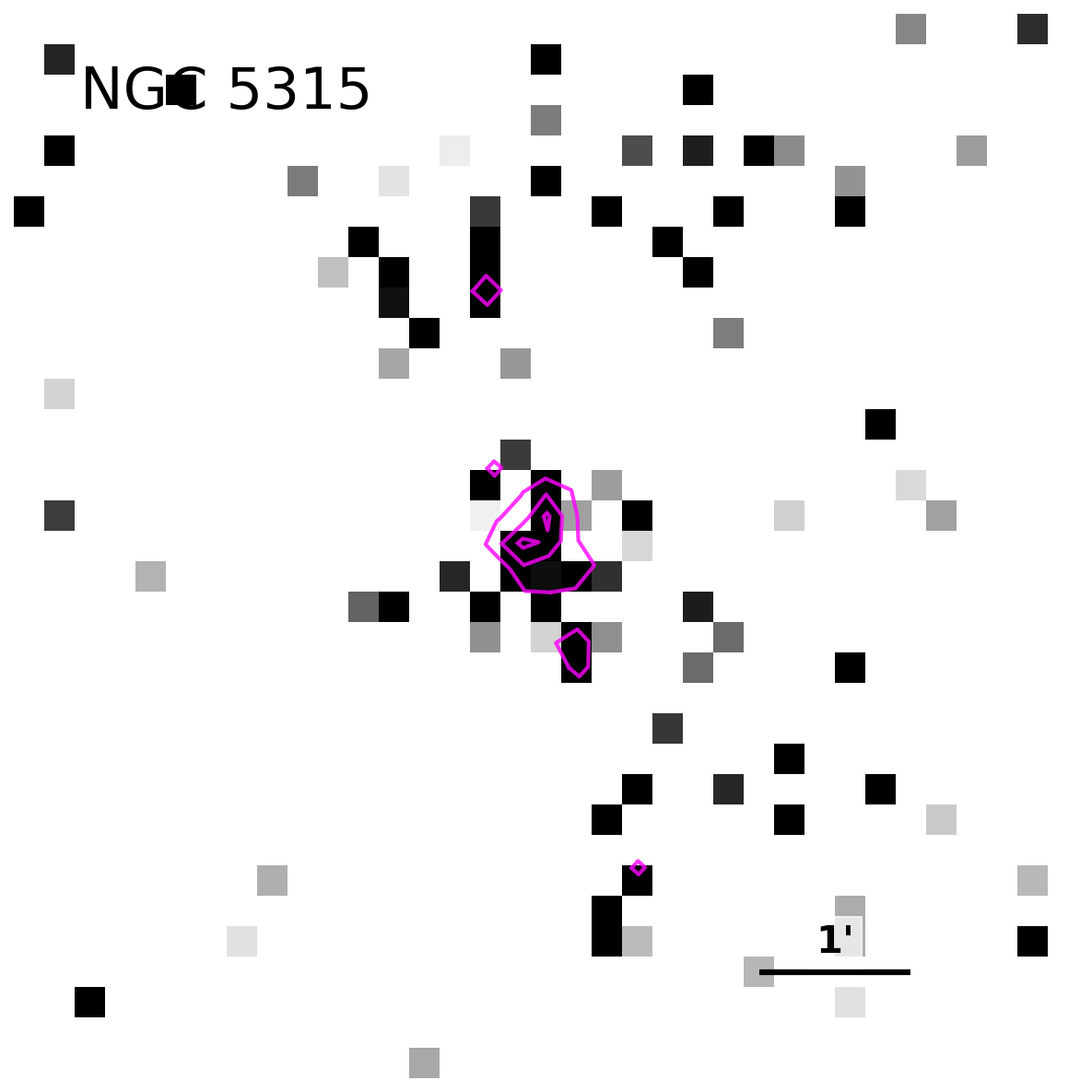}
\includegraphics[width=0.25\textwidth]{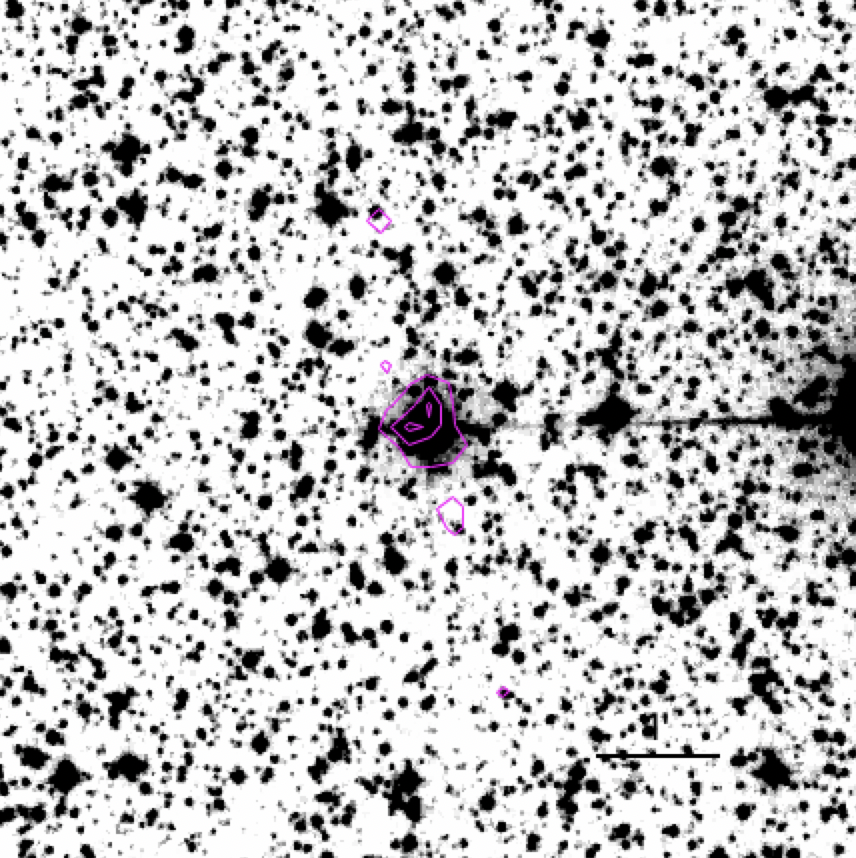}
\includegraphics[width=0.25\textwidth]{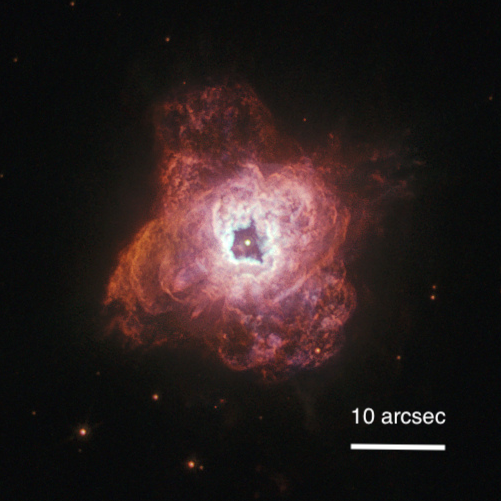} 
\includegraphics[width=0.25\textwidth]{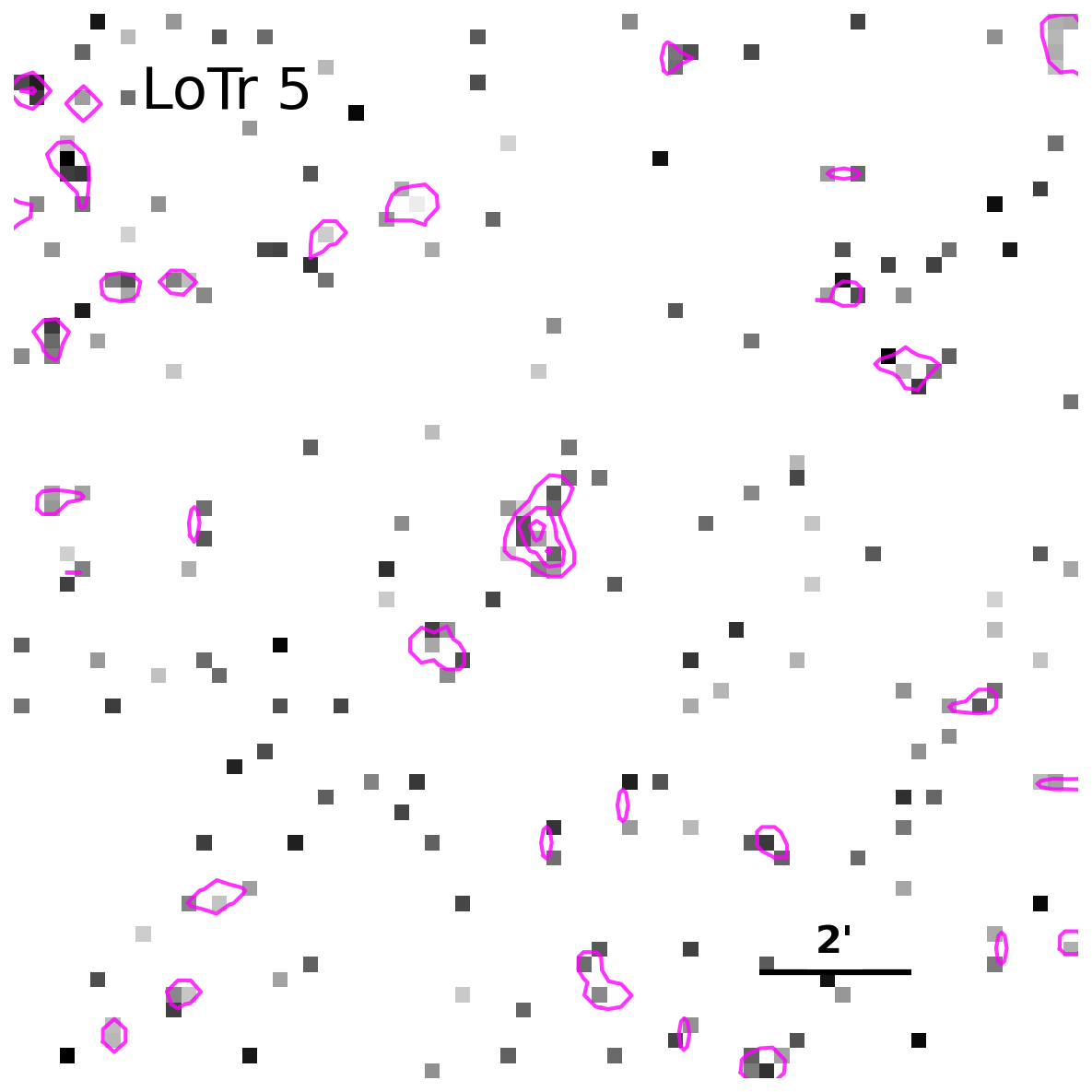}
\includegraphics[width=0.25\textwidth]{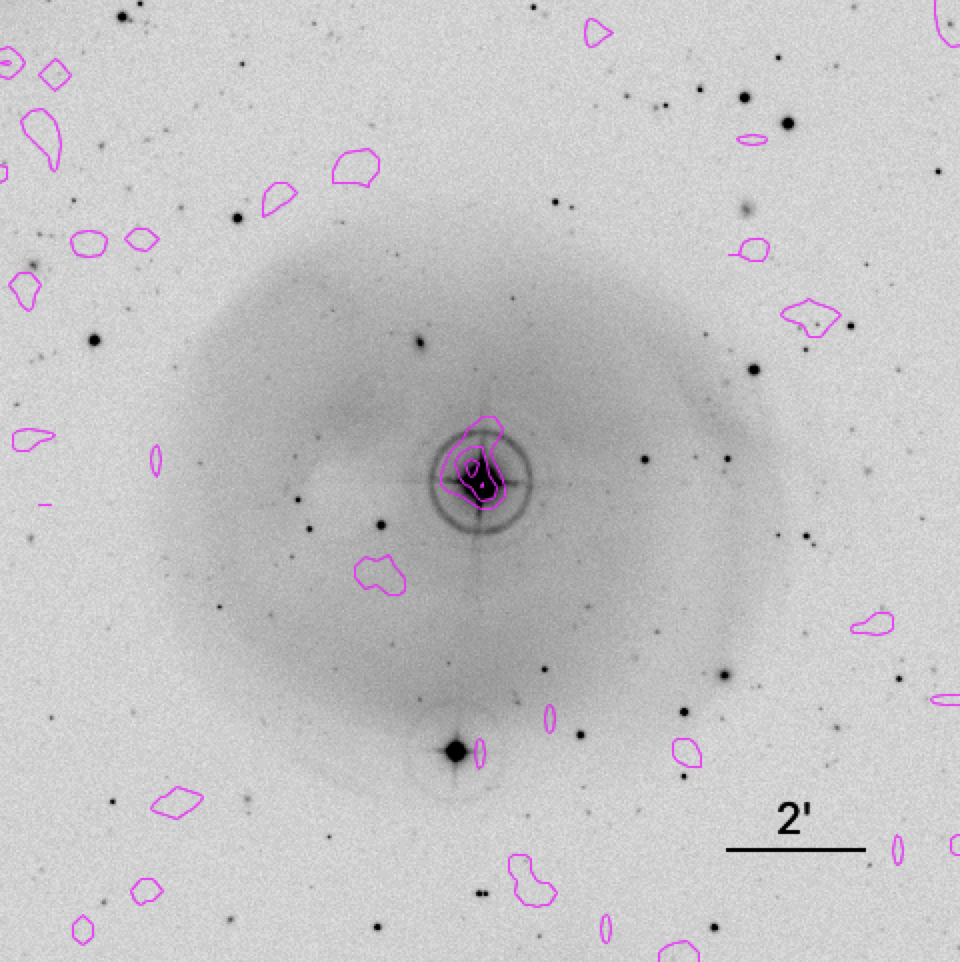}
\includegraphics[width=0.25\textwidth]{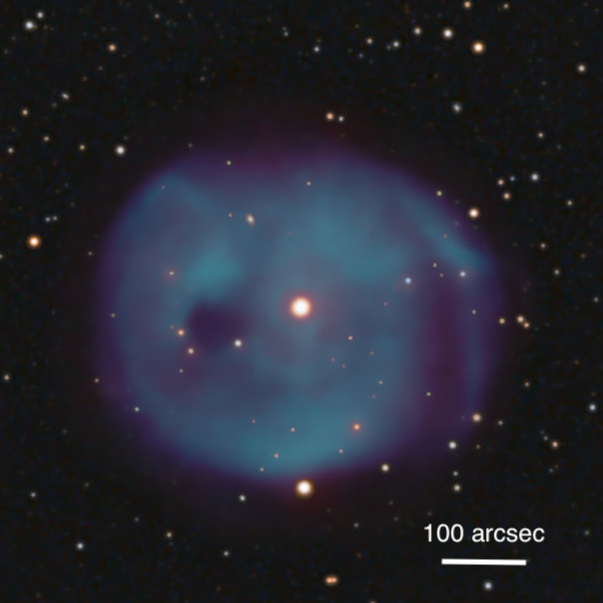} 
\caption{
{\it eROSITA} (left), B-band (middle) and colour-composite optical (right) images of the five already known X-ray PNe.
The X-ray and photographic B-band images (left and middle panels) have a similar $7\times7$ arcmin field of view(LoTr\,5 has $14\times14$ arcmin) and are 
overlaid by X-ray contours. NGC\,2392 and LOTR\,5's optical images are from J-PLUS.
North is up, east to the left. The final colour image is also NE to top left 
but is zoomed in to provide the PNe morphological detail, except for LoTr 5, which is already 
well resolved. The first and 4th colour images are from the HST (image credit NASA/HST), 
while the 4 other blue dominated images are from the very deep amateur narrow-band [OIII] dominated imagery of Peter Goodhew: https://www.imagingdeepspace.com/
}
\label{fig:pn_img_ok}
\end{figure}

\begin{figure}[t!]
\centering
\includegraphics[width=0.25\textwidth]{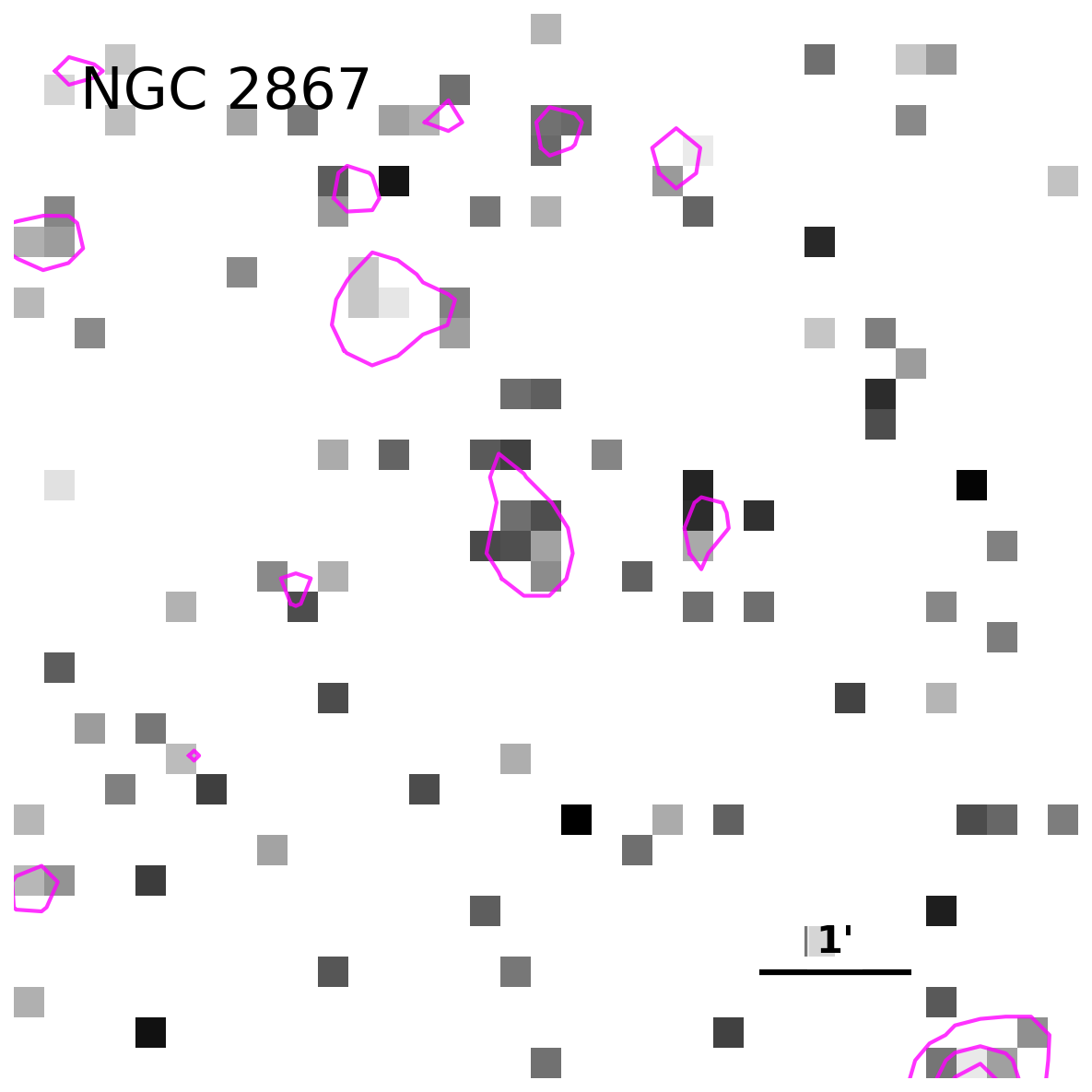}
\includegraphics[width=0.25\textwidth]{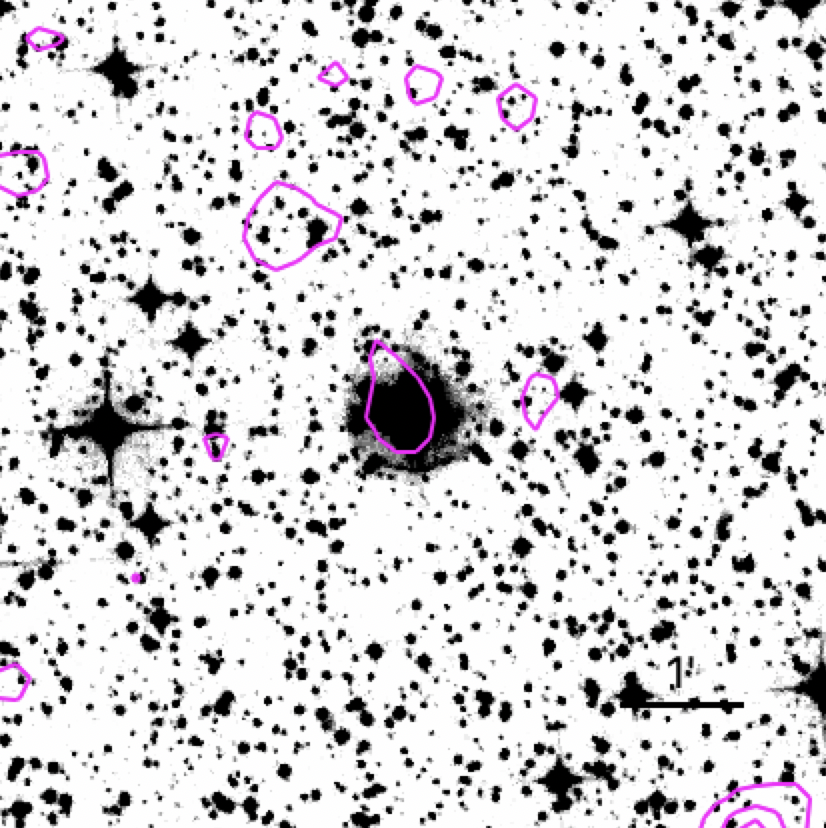}
\includegraphics[width=0.25\textwidth]{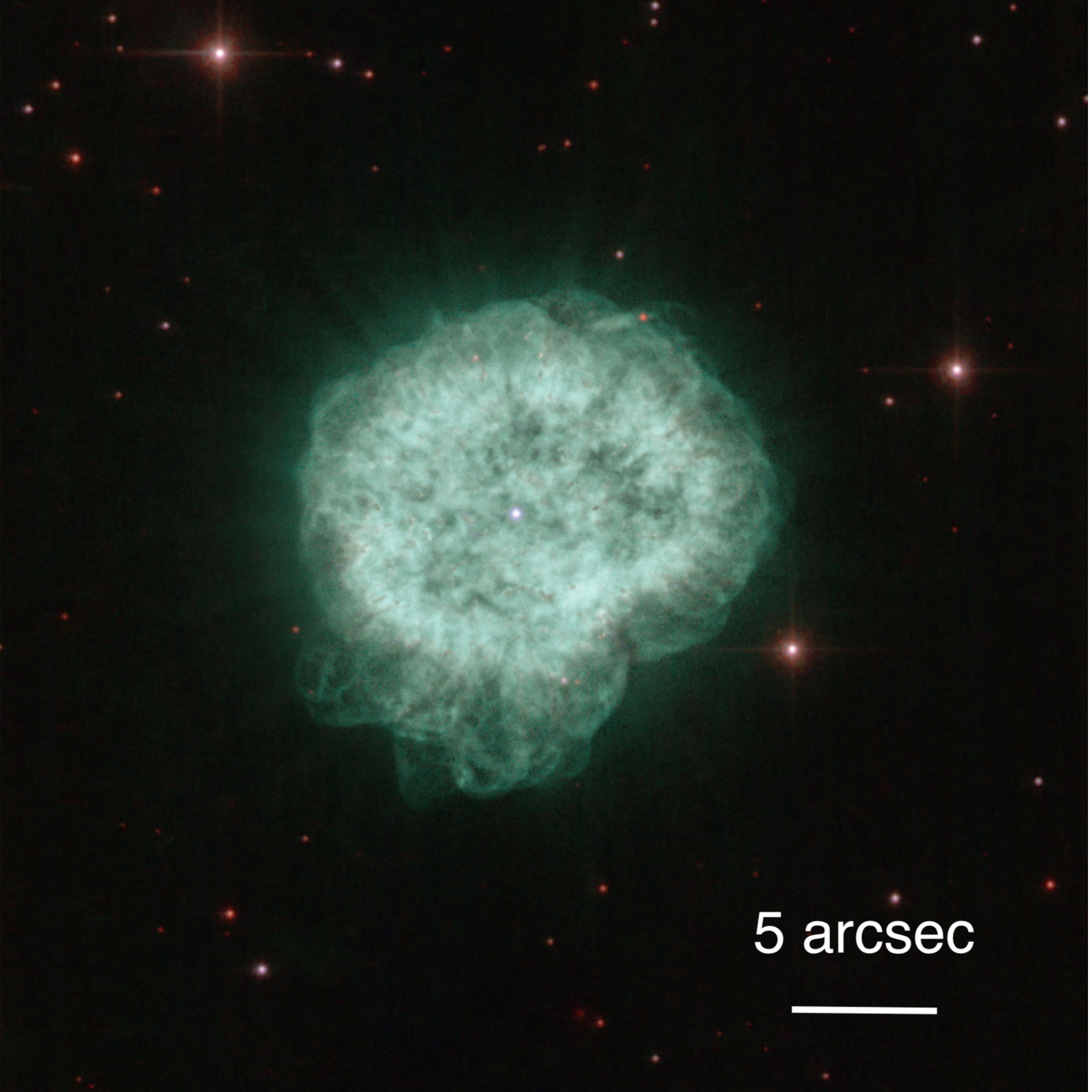} \\
\includegraphics[width=0.25\textwidth]{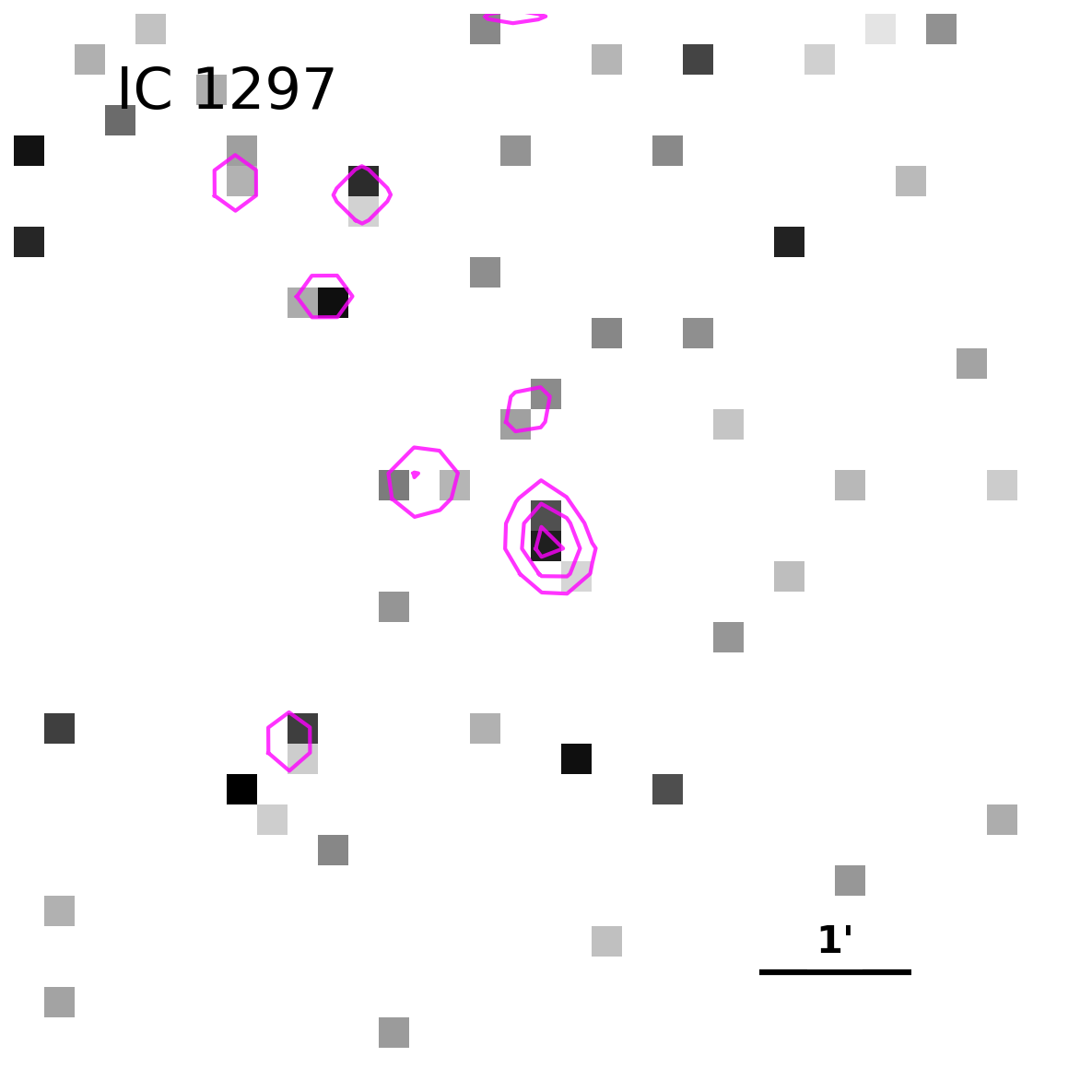}
\includegraphics[width=0.25\textwidth]{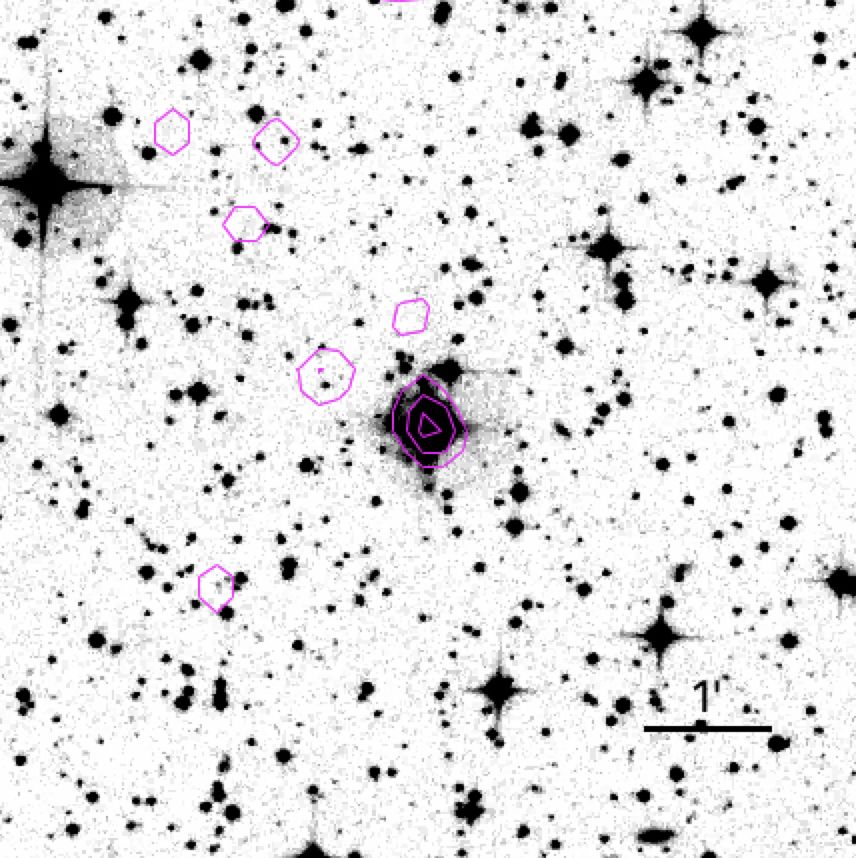}
\includegraphics[width=0.25\textwidth]{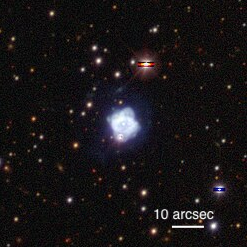} \\
\caption{
Same as Fig.~\ref{fig:pn_img_ok} for the two new X-ray PNe uncovered in eRASS1, with NGC\,2867  top and IC\,1297 bottom. 
The colour image for NGC\,2867 is from the HST (image credit NASA/HST) while that for IC\,1297 is from Legacy Surveys/D. Lang 
(Perimeter Institute) Filters: g, r, z.}
\label{fig:pn_img_new}
\end{figure}

\begin{figure}[t!]
\centering
\includegraphics[width=0.25\textwidth]{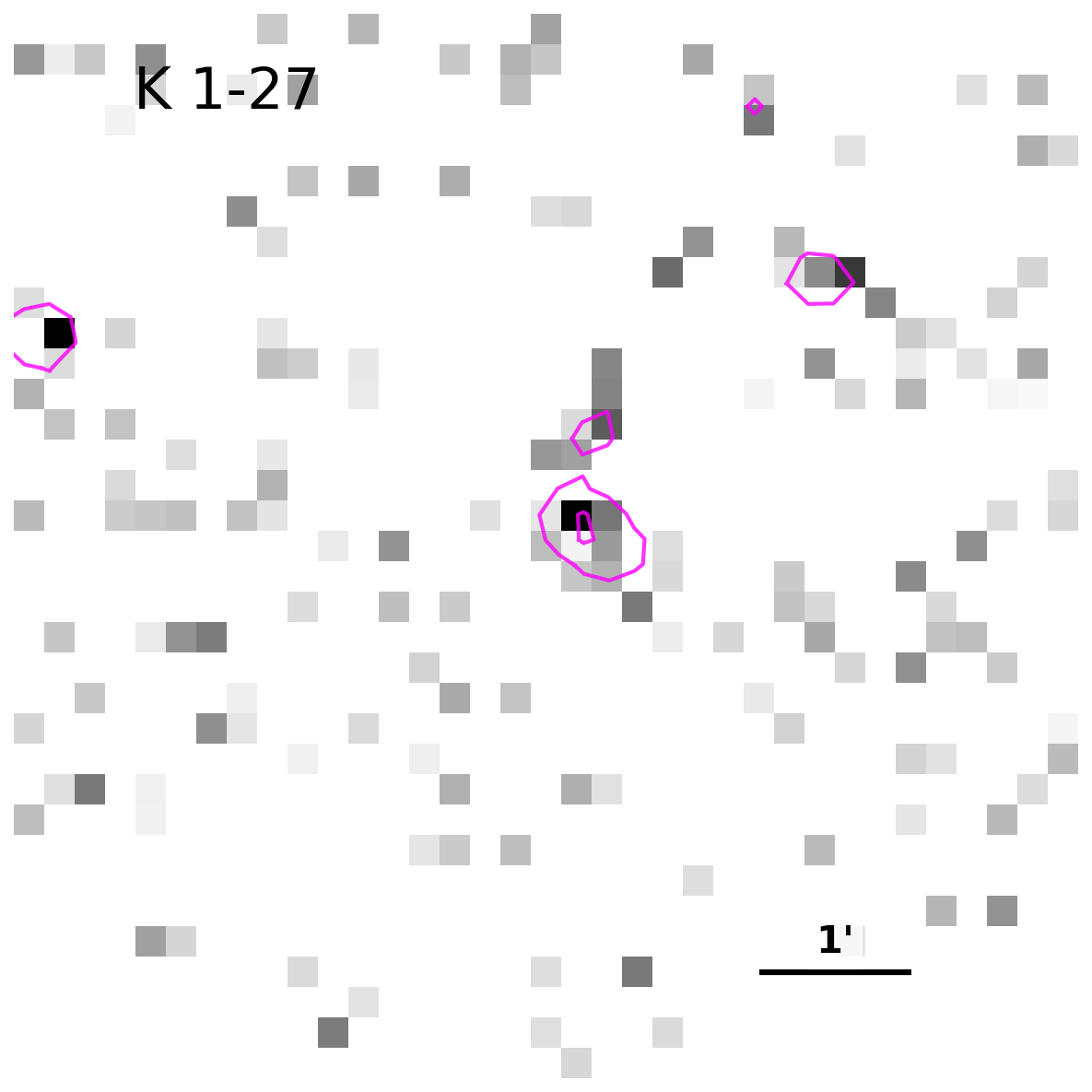}
\includegraphics[width=0.25\textwidth]{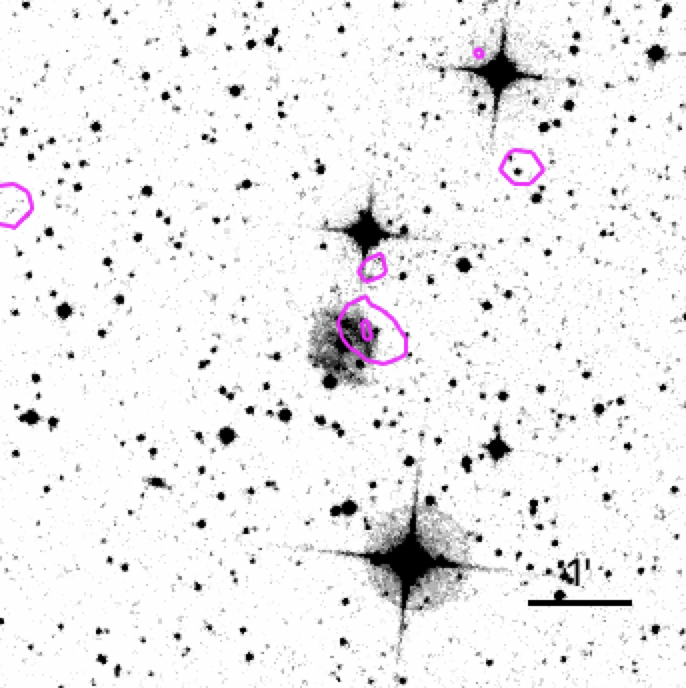}
\includegraphics[width=0.25\textwidth]{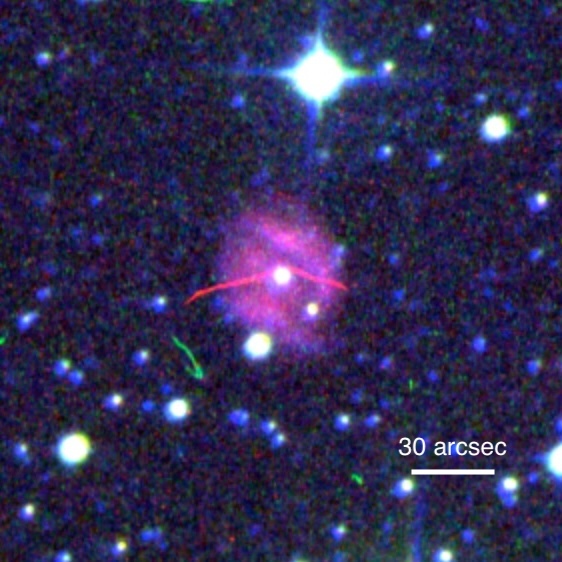}
\includegraphics[width=0.25\textwidth]{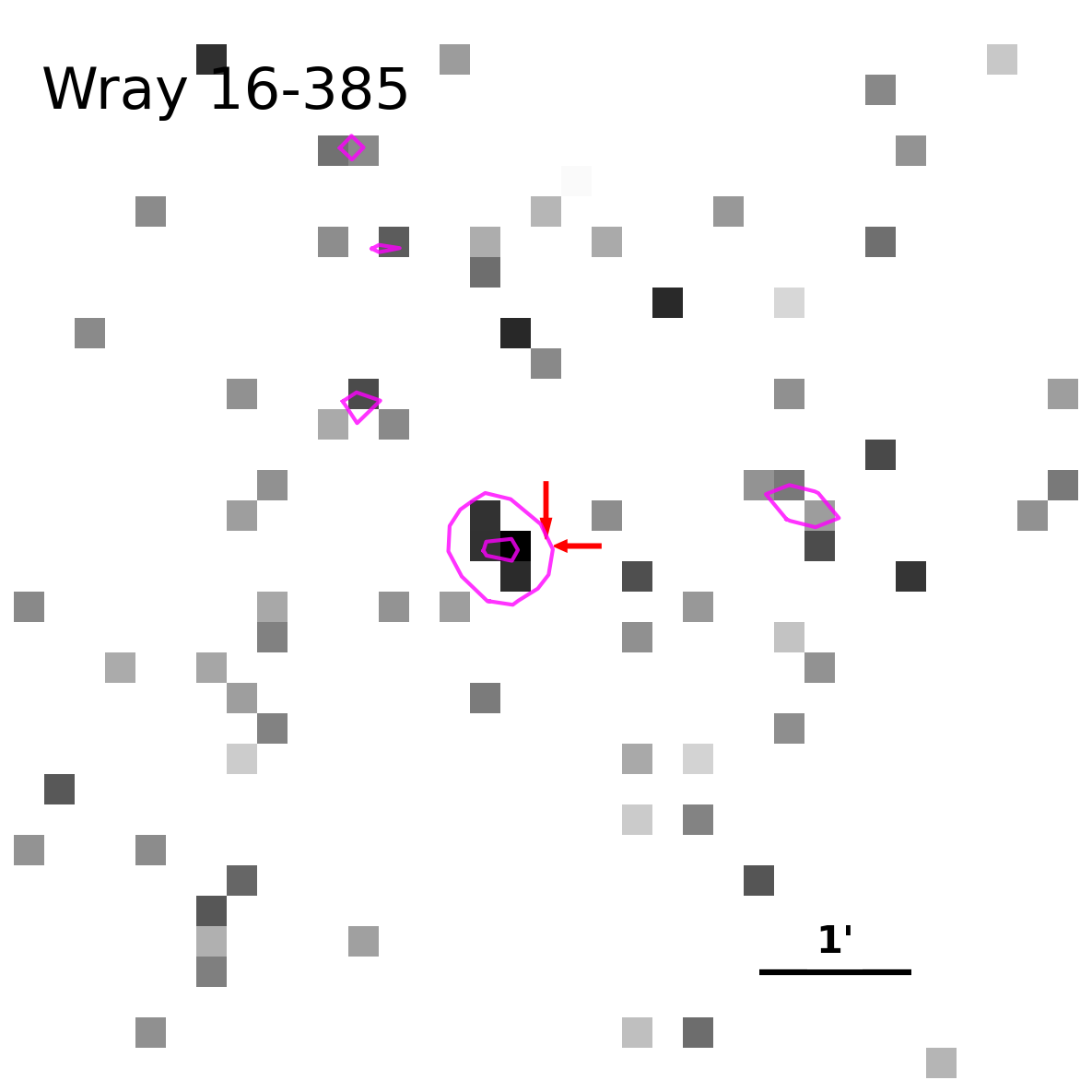}
\includegraphics[width=0.25\textwidth]{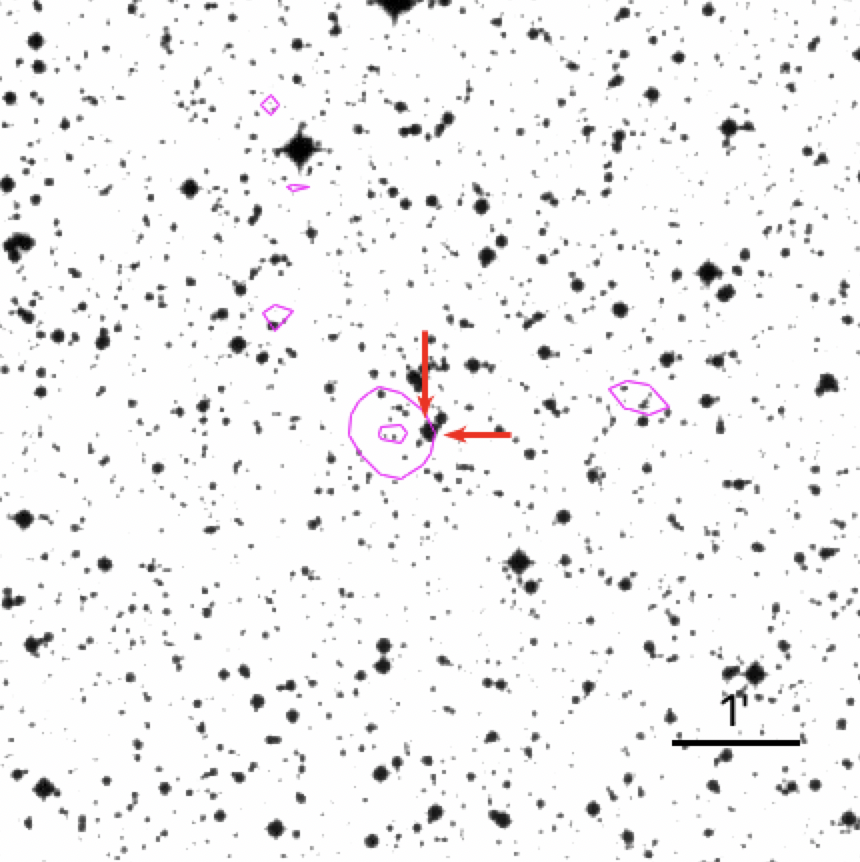}
\includegraphics[width=0.25\textwidth]{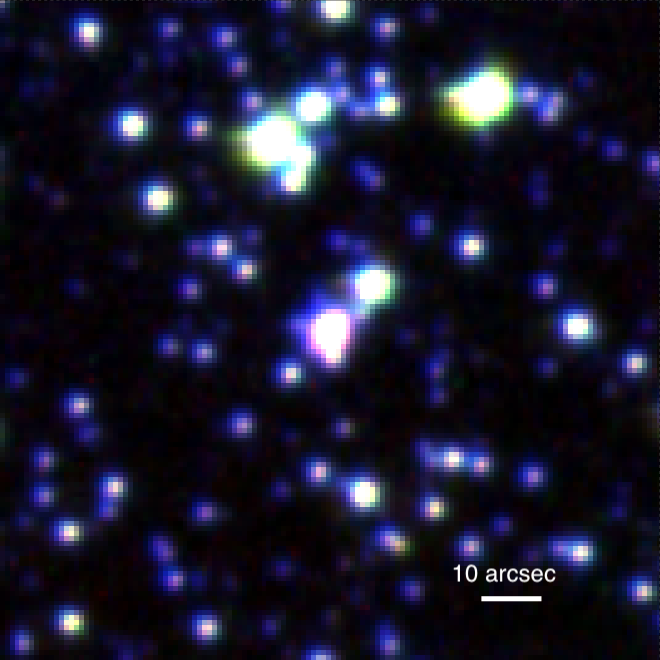}
\includegraphics[width=0.25\textwidth]{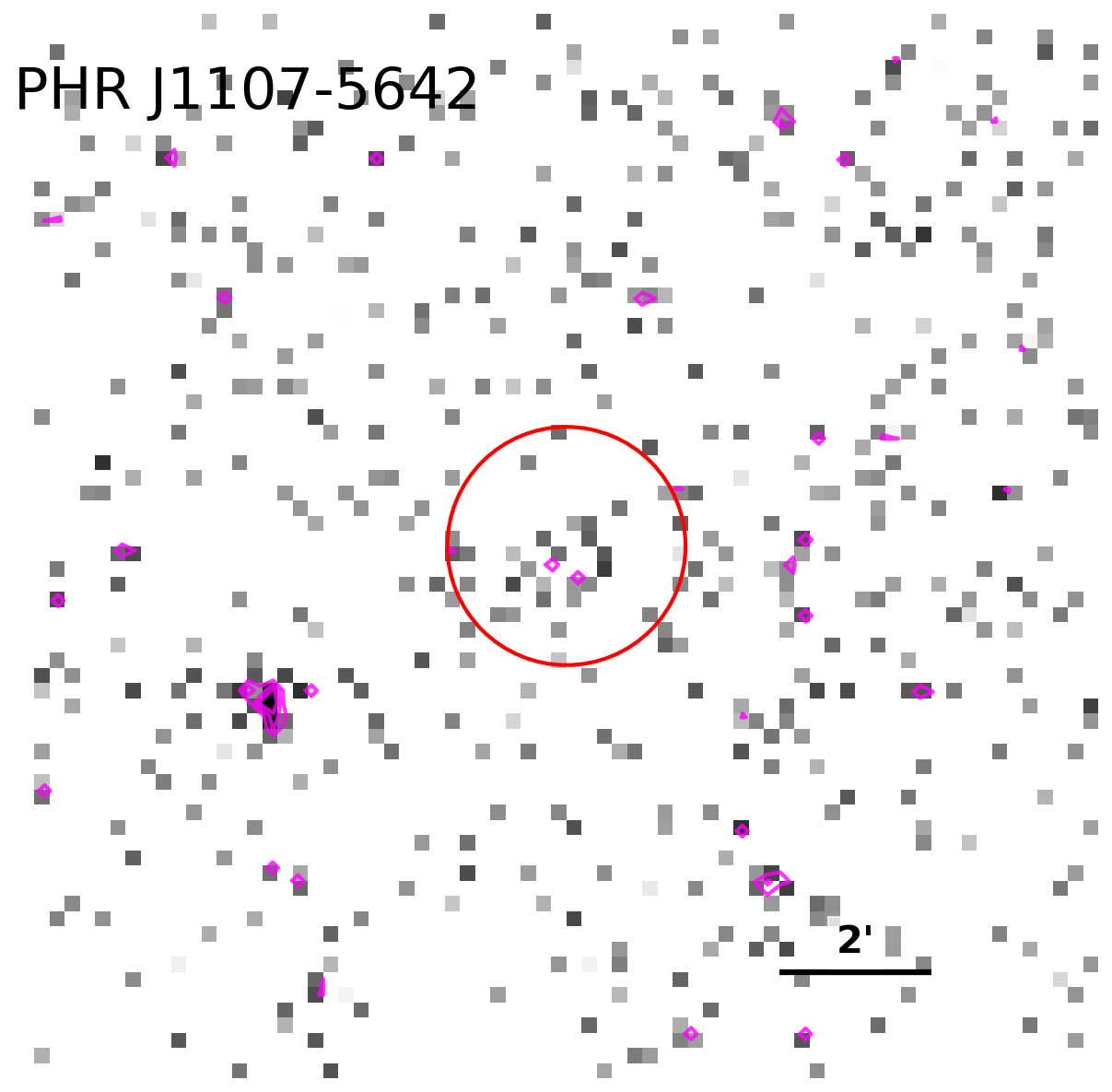}
\includegraphics[width=0.25\textwidth]{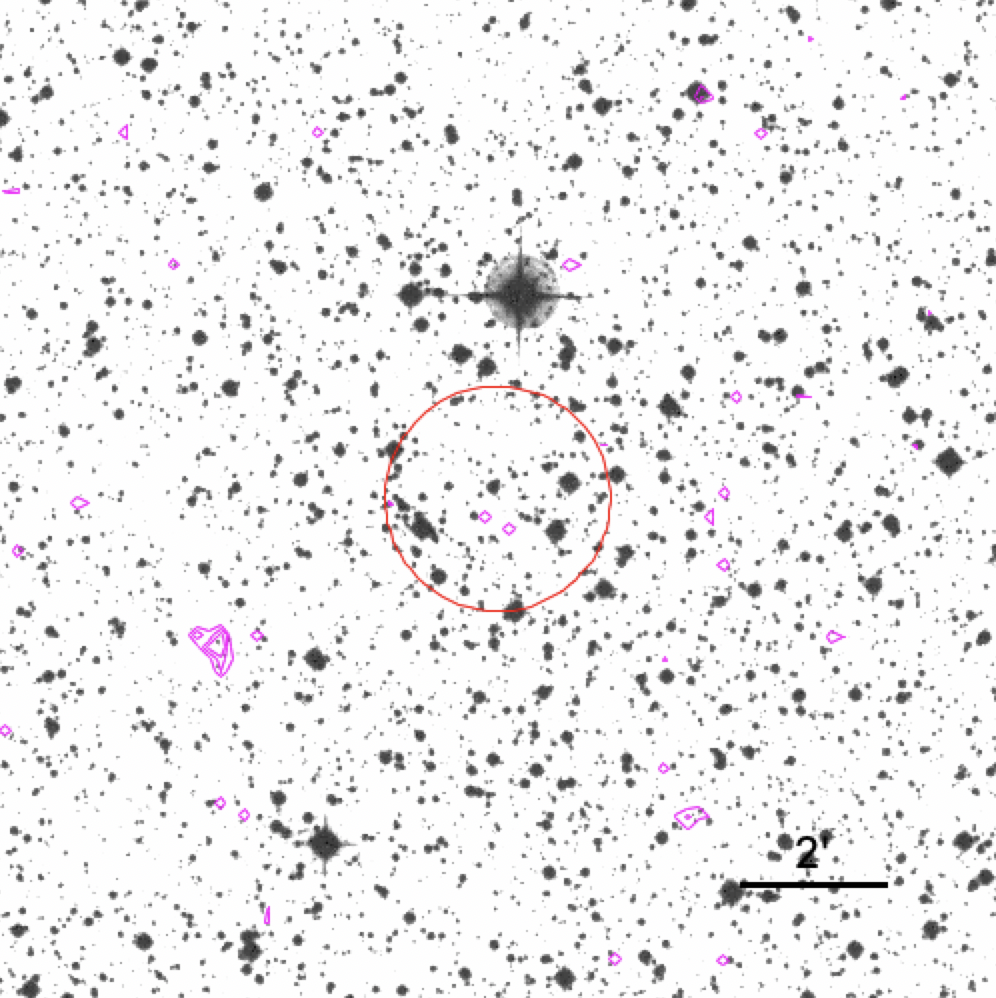}
\includegraphics[width=0.25\textwidth]{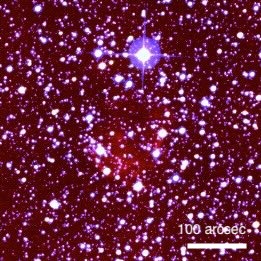}
\caption{
{\it eROSITA} (left) and B-band (middle) images of the three PNe 
with spurious eRASS1 counterparts overlaid by X-ray contours. The X-ray and B-band images have the same $7\times7$ arcmin field of view(PHR\,J1107$-$5642  has $14\times14$ arcmin).  North is up, east to the left. The final image in each case is an RGB colour combination from the SHS and SuperCOSMOS sky survey photographic data, with H$\alpha$ the red channel, SR the green channel, and B$_j$ the blue channel, all adjusted to best reveal the PN. The red arrows on the {\it eROSITA} and B-band images for Wray 16-385 show the location of the PN offset from the X-ray source
}
\label{fig:pn_img_pos}
\end{figure}

The quality of the association between these 10 PNe and their possible X-ray counterparts in the eRASS1 main source catalogue 
was then individually assessed.  {\it eROSITA} 0.2-2.0 keV X-ray images of these eRASS1 PN candidates were extracted from 
the event files and then compared to their optical images to make this assessment (Figs.~\ref{fig:pn_img_ok}, \ref{fig:pn_img_new} 
and \ref{fig:pn_img_pos}). We present 3 images for each X-ray candidate per row. 
The leftmost panels show the {\it eROSITA} X-ray photon detections. 
The pixel size of these images has been binned to 10~arcsec to properly sample the $\simeq$30~arcsec spatial resolution of {\it eROSITA}. 
The images are additionally overlaid with X-ray contours derived from images smoothed using a Gaussian profile with a 2-pixel kernel.
The images cover an area of 7$\times$7~arcminutes in and around the PN to show the X-ray detection in better context with 
the background, given the low count rates typical for all X-ray sources. In the middle panels we show, at the same angular size, 
the grey-scale, optical, narrow-band B-band image of the PNe taken from the SuperCOSMOS sky survey \citep{Parker2005} 
for all southern objects and JPLUS \citep{2019A&A...622A.176C} for the Northern hemisphere objects
so that the positional coincidence can be discerned. These images are overlaid with the same X-ray contours.  
Finally, as it is hard to see any structural detail in the optical PN images, given their generally small angular size 
and high surface brightness in the larger area 7$\times$7~arcminutes images, we provide in the rightmost panels zoomed in 
colour, combined narrow-band images of each PNe from the best available sources, including from HST and deep amateur 
imagery from Peter Goodhew . The interesting morphological structures and CSPN detections can now be properly appreciated 
in Figs.~\ref{fig:pn_img_ok}, \ref{fig:pn_img_new} and \ref{fig:pn_img_pos}. 

\begin{table*}
\centering
\caption{
X-ray-emitting true Galactic PNe now also detected in the eROSITA-DE eRASS1 catalog.
}
\label{tbl:pn_lst_ok}
\tiny
\setlength{\tabcolsep}{3pt} 
\renewcommand{\arraystretch}{1.1}
\begin{tabular}{lclcrccccc}
\hline
PNG &HASH & Common & eRASS1  & \multicolumn{1}{c}{Offset}   & \multicolumn{1}{c}{Position} & Nebular & Median & Count & Count Rate \\
      Name     &     \#        &       Name     & \multicolumn{1}{c}{IAU name} & \multicolumn{1}{c}{} & \multicolumn{1}{c}{1-$\sigma$} & diameter & energy & number \\
           &             &                 & \multicolumn{1}{c}{(J2000 RA/DEC)} & \multicolumn{1}{c}{(arcsec)} & \multicolumn{1}{c}{(arcsec)} & (arcsec)& {(KeV)} & & (s$^{-1}$) \\
           
\hline
PNG 197.8+17.3  &730 & NGC\,2392 & 1eRASS J072910.7$+$205444 &  2.4~~~~ & 4.9 &  46~~&0.48~ &   $7.1\pm2.8$ & $0.094\pm0.038$ \\
PNG 208.5$+$33.2& 742 & Abell\,30 & 1eRASS J084653.3$+$175248 &  3.0~~~~ & 6.5 & 127~~&0.40 &   $5.2\pm2.6$ & $0.071\pm0.036$ \\
PNG 261.0+32.0  & 824 & NGC\,3242 & 1eRASS J102446.3$-$183822 & 10.8~~~~ & 4.1 &  32~~&0.46 &   $6.2\pm2.7$ & $0.083\pm0.036$ \\
PNG 309.1$-$04.3& 955 & NGC\,5315 & 1eRASS J135357.4$-$663046 &  5.4~~~~ & 2.0 &  11~~&0.62 &  $43.2\pm7.1$ & $0.264\pm0.043$ \\
PNG 339.9+88.4  & 1082 & LoTr\,5   & 1eRASS J125533.9$+$255339 &  9.0~~~~ & 3.6 & 525~~&0.73 &  $17.9\pm4.7$ & $0.136\pm0.035$ \\
\hline
\end{tabular}
\end{table*}

\noindent
\begin{table*}
\centering
\caption{
Two true Galactic PNe with new X-ray detections in the eROSITA-DE eRASS1 catalog.
}
\label{tbl:pn_lst_new}
\tiny
\setlength{\tabcolsep}{3pt} 
\renewcommand{\arraystretch}{1.1}
\begin{tabular}{lclcrccccc}
\hline
PNG &HASH & Common & eRASS1  & \multicolumn{1}{c}{Offset}   & \multicolumn{1}{c}{Position} & Nebular & Median & Count & Count Rate \\
      Name     &     \#        &       Name     & \multicolumn{1}{c}{IAU name} & \multicolumn{1}{c}{} & \multicolumn{1}{c}{1-$\sigma$} & diameter & energy & number \\
           &             &                 & \multicolumn{1}{c}{(J2000 RA/DEC)} & \multicolumn{1}{c}{(arcsec)} & \multicolumn{1}{c}{(arcsec)} & (arcsec)& {(KeV)} & & (s$^{-1}$) \\
           \hline
PNG 278.1$-$05.9 & 858  & NGC\,2867 & 1eRASS J092125.6$-$581843 &  3.6~~~~ & 3.6 &        14.4    & 0.55  &    $8.7\pm3.3$ & $0.032\pm0.012$ \\
PNG 358.3$-$21.6 & 1267 & IC\,1297  & 1eRASS J191723.2$-$393639 &  6.6~~~~ & 3.7 &         10.8   &0.44  &   $7.3\pm4.0$ & $0.091\pm0.038$ \\
\hline
\end{tabular}
\end{table*}

\noindent
\begin{table*}
\centering
\caption{Three true Galactic PN with spurious X-ray counterparts in the eROSITA-DE eRASS1 catalog.}
\label{tbl:pn_lst_pos}
\tiny
\setlength{\tabcolsep}{3pt} 
\renewcommand{\arraystretch}{1.1} 
\begin{tabular}{lclcrcccccc}
\hline
PNG &HASH & Common & eRASS1  & \multicolumn{1}{c}{Offset}   & \multicolumn{1}{c}{Position} & Nebular & Median & Count & Count Rate \\
      Name     &     \#        &       Name     & \multicolumn{1}{c}{IAU name} & \multicolumn{1}{c}{} & \multicolumn{1}{c}{1-$\sigma$} & diameter & energy & number \\
           &             &                 & \multicolumn{1}{c}{(J2000 RA/DEC)} & \multicolumn{1}{c}{(arcsec)} & \multicolumn{1}{c}{(arcsec)} & (arcsec)& {(KeV)} & & (s$^{-1}$) \\

\hline
PNG 286.8$-$29.5 & 888 & K\,1-27           & 1eRASS J055657.9$-$754015 & 17.4 & 3.7 & 61 &1.01& $12.5\pm4.8$ & $0.021\pm0.008$ \\
PNG 351.7$-$10.9 & 1157 & Wray\,16-385      & 1eRASS J181254.3$-$413029 & 16.2 & 2.7 & 8 &0.65 &\ $15.4\pm4.3$ & $0.182\pm0.501$  \\
PNG 289.0$-$03.3& 2672 & PHR\,J1107$-$5642 & 1eRASS J110744.1$-$564240 & 16.2 & 5.6 & 188 & 0.91&$8.4\pm3.8$ & $0.035\pm0.015$\\
\hline
\end{tabular}
\end{table*}

\begin{figure}[htb!]
\centering
\includegraphics[width=0.75\textwidth]{ 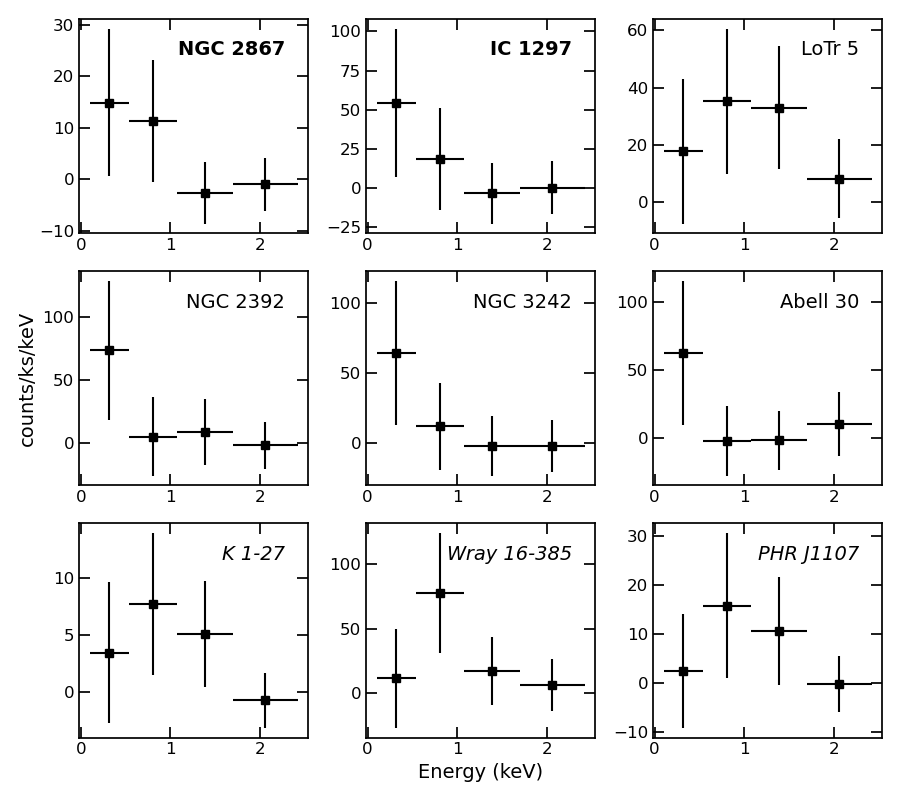}
\caption{
{\it eROSITA} background-subtracted spectra of the X-ray counterparts of PNe in eRASS1 with limited count number. The peak at low photon energy distributions for the 5 confirmed X-ray emitting PNe compared to the 3 rejected candidates K 1-27, Wray 16-385 and PHR~J1107-5642 is clear.
}
\label{fig:spec_all}
\end{figure}

After inspecting those images, any PN with a positional coincidence between the HASH and the eRASS1 counterpart, 
accounting both for the eRASS1 positional uncertainty and the PN angular extent, and with a count rate at least 
twice the count rate uncertainty were considered bona-fide X-ray PNe in the {\it eROSITA} eRASS1 catalogue.  
These include five PNe already known to be X-ray-emitters (Tab.~\ref{tbl:pn_lst_ok}, Fig.~\ref{fig:pn_img_ok}) and 
another two newly identified X-ray emitting PNe, IC\,1297 and NGC\,2867 (Tab.~\ref{tbl:pn_lst_new}, Fig.~\ref{fig:pn_img_new}).  
Another three not fulfilling the above criteria or resulting 
in unconvincing matches after the examination of their optical and X-ray images were found. This includes K\,1-27 that was 
previously identified as a PN exhibiting X-ray emission \citep{Rauch1994}. The two other potential X-ray emitting PNe considered 
unlikely X-ray PNe in the {\it eROSITA} eRASS1 catalogue after vetting here (Tab.~\ref{tbl:pn_lst_pos}, Fig.~\ref{fig:pn_img_pos}), 
are Wray\,16-385, and PHR\,J1107$-$5642. The {\it eROSITA} background-subtracted X-ray spectra of these PNe are presented in 
Figs.~\ref{fig:spec_all} and \ref{fig:spec_NGC5315}. 

A short description of the PNe and the spatial and spectral properties of their associated X-ray emission uncovered by 
this cross-matching process is provided below for each case referred to above.  This includes PNe previously known 
to have X-ray detections and new candidates uncovered by this work, as well as the deduced spurious eRASS1 counterparts of PNe.

\subsection{Known X-ray Galactic PNe with eROSITA eRASS1 counterparts}
In this section, we examine the known X-ray emitting PNe that fall within the currently accessible survey footprint of {\it eROSITA}.

\subsubsection{NGC\,2392, the Eskimo Nebula}

NGC\,2392 is the famous Eskimo Nebula, with a double-shell morphology in the optical. It has been imaged on multiple occasions and 
with a significant associated literature, e.g. \citep[e.g.,][]{1985ApJ...295L..17G,2005A&A...430L..69G, 2012ApJ...761..172G}.
Its X-ray emission was only tentatively detected in the ROSAT PSCP observations \citep{Guerrero2000ROSAT} with a 
count rate of $0.0021\pm0.0013$ s$^{-1}$. 
The diffuse X-ray emission is confined within its innermost shell as confirmed by XMM-Newton \citep{Guerrero2005XMMEskimo}. 
The diffuse X-ray emission from NGC\,2392 can be described by a hot plasma with a temperature 
$\simeq2\times10^6K$ and an X-ray flux in the 0.2–2.5 keV energy range of $(6\pm1)\times10^{-14}$ erg~cm$^{-2}$~s$^{-1}$.  
At the Gaia distance of $1830\pm90$ pc for its CSPN, the intrinsic X-ray luminosity would be $(4.1\pm1.6)\times10^{31}$ erg~s$^{-1}$. 
The spatial distribution of the X-ray emission in the 0.2-0.65 keV band was suggested to align along the fast collimated outflow, 
but Chandra X-ray observations at much improved spatial resolution do not confirm it.  Instead, it revealed the presence 
of a point-source of hard, up to 3 keV X-ray emission at the central star position 
of NGC\,2392, see \citep[see][]{Ruiz_2013NGC5315_eskimo, ChanPlaNS1}.  
Once the emission from this source is excised, the X-ray flux and intrinsic luminosity are found to 
be $3.9\times10^{-14}$ erg~cm$^{-2}$~s$^{-1}$ and $2.6\times10^{31}$ erg~s$^{-1}$, respectively. 
The peculiar, hard X-ray emission of its CSPN is attributed to local scale accretion from a companion, 
whereas the origin of the diffuse X-ray emission is enigmatic, as the CSPN wind cannot provide the mechanical 
energy that the observed plasma temperature and luminosity require \citep{Guerrero2019eskimo, NGC2392binaryEnvelop}. 
Furthermore, the diffuse emission is confined within a hot "flocullent" bubble which appears to be a strong morphological 
trait for all other Galactic PNe with diffuse X-ray emission.  A single-lined spectroscopic binary with 1.9-day 
period in NGC\,2392 was confirmed by \citep{2019PASA...36...18M} from a careful, ECHELLE spectrograph-based, radial velocity 
study of the CSPN.

The X-ray emission detected by {\it eROSITA} perfectly matches the position of NGC\,2392, with most X-ray photons enclosed 
within the overall inner nebular shell. The low count number, $\simeq$7, is insufficient for spectral analyses 
(Fig.~\ref{fig:spec_all}). The median energy of these events is 0.482 keV, peaking below 0.5 keV.  
The location of NGC\,2392 on Fig.~\ref{fig:medeplot} implies spectral properties indicative of a plasma with 
temperature of $(1-2)\times10^6$~K, which is consistent with the diffuse emission source of NGC\,2392 detected 
by XMM-Newton and Chandra.  


\subsubsection{Abell\,30}
Abell\,30 is another very unusual, so-called "born again" PN \citep[e.g.][]{1983ApJ...266..298J, Wesson+2005} with a high 
abundance discrepancy factor shown by \citet{Wesson+2018} to be a strong indicator of CSPN binarity and recently confirmed as such \citep{Jacoby+2020}. 
ROSAT first detected diffuse X-ray emission with plasma temperatures of $(2-4.5)\times10^5K$, fitted by an optically-thin plasma 
emission model. The X-ray emission is very soft, with almost all source counts at energies below 0.4 keV~\citep{Chu1995A30}.
The comparison with HST WFPC2 imagery of this PN revealed the significant spatial 
correspondence between the X-ray and optical emission centroids for both a bipolar pair of knots and a clumpy, expanding disk 
\citep{Chu1997A30}. The diffuse X-ray emission of Abell\,30 has been suggested to be generated by the fast stellar wind 
interactions with the hydrogen-poor ejecta of the reborn process, either by stellar wind ablation in high electron density 
environments ($~\thicksim 1000$~cm$^{-3}$) or by charge-exchange reactions with neutral material \citep{Guerrero2012A30}.
Analyses of XMM-Newton and Chandra spectra imply a plasma temperature of $0.79\times10^6$~K. 
The observed flux in $0.2-1.5$~keV energy range is $(2.8\pm0.9)\times10^{-14}$~erg~cm$^{-2}$~s$^{-1}$, and the intrinsic 
luminosity at the Gaia distance of $2210\pm160$ pc is $\sim 1.3~\times~10^{31}$~erg~$s^{-1}$.

The X-ray emission detected by {\it eROSITA} perfectly matches the position of Abell\,30, with most X-ray photons 
around the central star and H-deficient knots. The low X-ray photon count number, $\simeq$5, is again insufficient for 
spectral analyses. Their median energy is 0.40~keV and their X-ray spectrum is soft, peaking below 0.5~keV (Fig.~\ref{fig:spec_all}). 
The temperature implied by a typical plasma emission model is low, $\simeq 1\times10^6$~K (Fig.~\ref{fig:medeplot}), 
but higher than that derived by \citet{Guerrero2012A30}. The low count number and peculiarities of the X-ray emission 
from Abell\,30 and the absorption component certainly account for this discrepancy.

\subsubsection{NGC\,3242, the Ghost of Jupiter Nebula}
NGC\,3242 is another well-known, iconic, multiple shell PN \citep[e.g.][]{1985ApJ...294..193B, 2025A&A...697A.227K}, 
discovered by William Herschel in 1785. It is at a distance of $1340\pm90$~pc based on its CSPN Gaia detection.  
Diffuse X-ray emission within its innermost shell was first detected by XMM-Newton \citep{Ruiz2011NGC3242}.  
The X-ray emission is soft, with most photons with energies below 1.0 keV and very little emission at higher energies.  
It can be described by a thin-plasma emission model at a temperature of $\sim2.35\times10^6K$, with an observed X-ray flux of $4\times10^{-14}$ erg~cm${-2}$~s$^{-1}$ and an intrinsic X-ray luminosity in the 0.4-2.0 keV band of $4\times10^{31}$ erg~s$^{-1}$.  
In this sense, the detection of these soft X-ray photons is favored by the low extinction towards NGC\,3242, with a hydrogen 
column density $N_H=5\times10^{20} cm^{-2}$ according to the logarithmic extinction coefficient of $c(H_\beta)=0.12$ 
derived by \citet{Pottasch2008NGC3242extiction}.  The diffuse X-ray emission from NGC\,3242 was confirmed to be confined within 
its innermost shell, which can then be described as a hot bubble powered by the stellar wind, by Chandra observations with superior 
spatial resolution \citep{ChanPlaNS1}.  There is no detectable X-ray emission from the central star. 

With a similar PSF to XMM-Newton, the X-ray emission detected by {\it eROSITA} perfectly matches the position of NGC\,3242, 
with most X-ray photons enclosed within the innermost shell. The count number, $\simeq6$, is once again insufficient for 
spectral analyses, but the spectral shape in Fig.~\ref{fig:spec_all} 
is consistent with the soft X-ray emission from this PN.  
The median energy of these events, $\simeq$0.46 keV, and low extinction suggest a plasma temperature in the range $(1-2)\times10^6$~K (Fig.~\ref{fig:medeplot})––v.

\subsubsection{NGC\,5315}

NGC\,5315 is another reasonably well-known, high surface brightness PN \citep[e.g.][]{2004ApJS..150..431P, 2002A&A...393..285P}. 
Its X-ray emission was serendipitously detected in Chandra observations (ObsID\,4480) that were actually targeting  another 
PN, Hen\,2-99 which ironically itself was not detected.  The X-ray spectrum of NGC\,5315 is consistent with emission from a 
hot plasma at $T_X\sim 2.5\times10^6$~K with enhanced Ne abundances \citep{Kastner_2008NGC5315}.  
The observed X-ray flux is $\simeq1\times10^{-13}$ erg~cm$^{-2}$~s$^{-1}$, for an intrinsic X-ray luminosity at the Gaia distance\footnote{\citet{Kastner_2008NGC5315} adopted a distance of 1.5 kpc, resulting in a larger estimate of the X-ray luminosity of 
$2.5\times10^{32}$ erg~$s^{-1}$.} of $4\times10^{31}$ erg~$s^{-1}$ in the $0.3-2.0$ keV energy band \citep{Kastner_2008NGC5315} 
of $960\pm190$ pc. 
The X-ray-luminous hot bubble presents emission in the soft 
band with the extinction of $c(H_{\beta})=0.54$, and column density of $N_H=2.2\times10^{21}~cm^{-2}$~\citep{Pottasch2002NGC5315}.

\begin{figure}[htb!]
\centering
\includegraphics[width=0.8\textwidth]{ 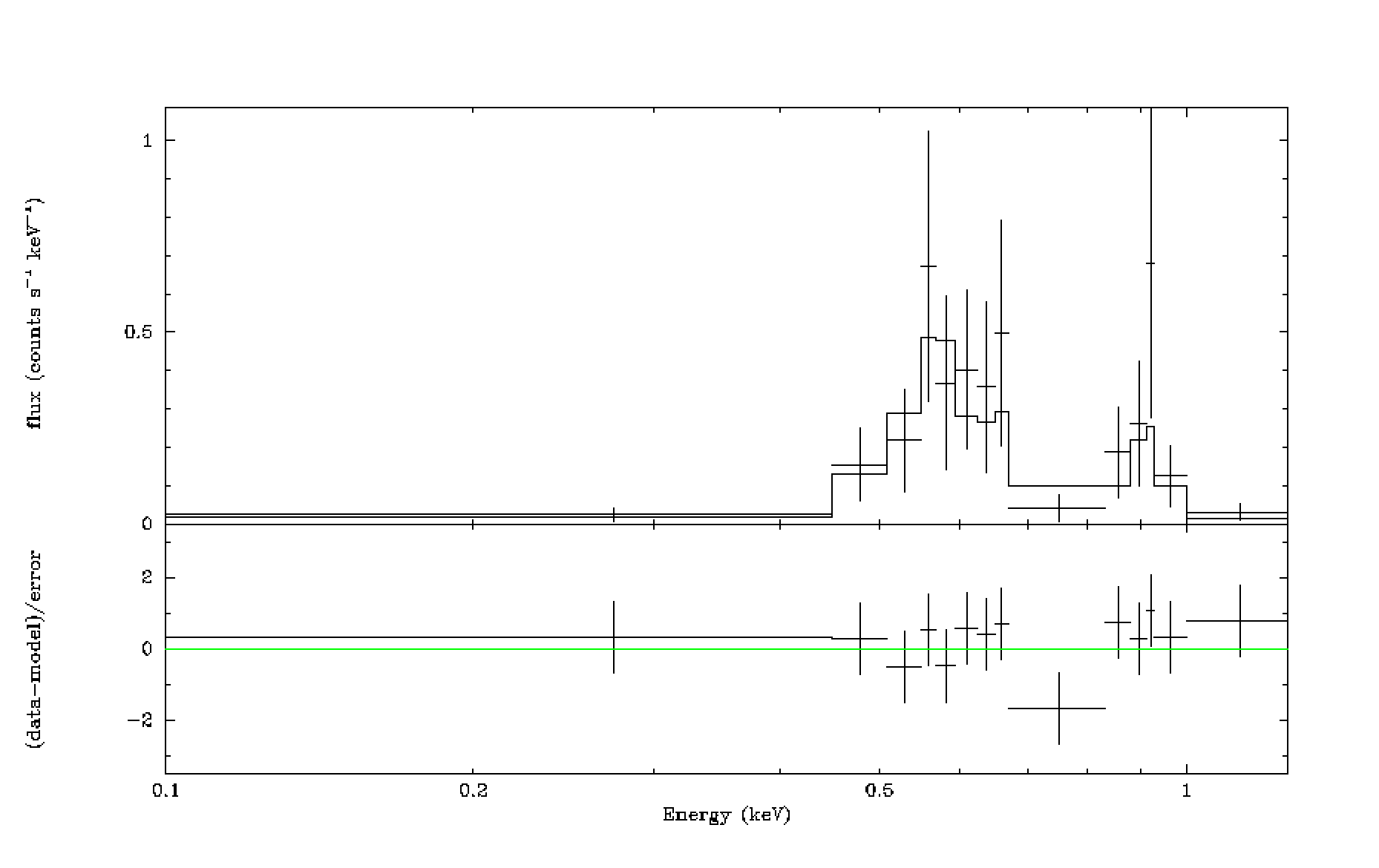}
\caption{
(top panel) {\it eROSITA} background-subtracted X-ray spectrum of NGC\,5315 (dots) overlaid with the best-fit model (histogram). 
(bottom) Residuals of the best-fit model. 
}
\label{fig:spec_NGC5315}
\end{figure}

The X-ray emission detected by {\it eROSITA} perfectly matches the position of NGC\,5315.
The count number, $\simeq$43, is sufficient for spectral analyses (Fig.~\ref{fig:spec_NGC5315}).  
The spectrum shows a prominent Ne~{\sc ix} line at $\approx$0.9 keV, which causes the increase of the median energy of 
the events, $\simeq$0.62 keV, over that of the previously discussed PNe.  
Adopting the same Ne and Fe abundances to be 4.3 and 0.6 times solar, respectively, derived by \citet{Kastner_2008NGC5315}, 
the best-fit parameters are a hydrogen column density $N_{\rm H} = 2.1\times10^{21}$ cm$^{-2}$, a plasma temperature 
of $1.9\times10^6$~K (see also Fig.~\ref{fig:medeplot}) and 
an observed flux in the 0.1-2.0 keV range of $(1.7\pm0.3)\times10^{-13}$ erg~cm$^{-2}$~s$^{-1}$.  
The intrinsic X-ray luminosity would be $\approx2.3\times10^{32}$ erg~s$^{-1}$.

\subsubsection{LoTr\,5}

LoTr\,5 is an extended ($\sim$7\arcmin$\times$8\arcmin\ in size) nebula of low surface brightness but with 
a bright CSPN. It was first reported detected in X-rays by ROSAT \citep{1992A&A...264..623K} and has been 
reasonably well studied \citep[e.g.][]{1997A&A...320..913T, 2017A&A...600L...9J}
It has been described to have a bipolar structure oriented almost along the line of sight \citep{Graham+2004}.  
It is located near RXJ\,1256.0+2556, a fossil galaxy group of high X-ray luminosity \citep{Khosroshahi+2007}.  
As a result, LoTr\,5 was serendipitously observed by both Chandra and XMM-Newton, thus providing new, valuable X-ray data. 
Its low density and large angular size imply an evolved nebula, unusual for the X-ray emitting hot bubble of a PN.  
A thorough analysis of these X-ray data actually finds that there is no diffuse X-ray emission from a hot bubble \citep{Montez+2010}, 
but it instead arises from its central star.  
This is the known long-period binary system IN\,Com \citep[first suggested by][]{1985A&A...151L..13A} 
formed by a hot, 150,000~K, CSPN and a rapidly-rotating G5~III companion\footnote{The system has 
been proposed to include an additional tertiary component to the giant component \citep{Jasniewicz+1987,Aller+2018}, 
which might be responsible for its unusual rapid rotation.}. Its X-ray emission can be described using a 
two-temperature optically-thin plasma model of $7.5\times10^6$ and (26-41)$\times10^6$ K, respectively \citep{Montez+2010}.  
The observed X-ray flux seemed to vary among the XMM-Newton and Chandra observations from $9\times10^{-14}$ 
to $2.3\times10^{-13}$ erg~cm$^{-2}$~s$^{-1}$, respectively, indicating possible intense activity.  
Accordingly, the intrinsic X-ray luminosity at the Gaia distance of $455\pm10$ pc ranges from 
$2.2\times10^{30}$ to $4.4\times10^{30}$ erg~s$^{-1}$, respectively.  

The new {\it eROSITA} detection confirms the association of the X-ray emission with the central star of LoTr\,5. 
The spectral shape is quite flat and indicates a hard spectrum, with a median energy of 0.73 keV, higher than 
the typical value associated with hot bubbles in PNe (Fig.~\ref{fig:medeplot}), but consistent with the hard 
X-ray emission from the G5~III component of IN\,Com. The background-subtracted spectrum has 15 counts. 
Adopting the two thin thermal plasma models, the plasma temperatures are simulated as 0.034 and 2.5 keV, 
respectively (with a reduced $\chi^2$ of 1.23). The flux is estimated to be $1.3\times 10^{-13}$~erg~cm$^{-2}$~s$^{-1}$. 
The X-ray emission level matches those of previous archival spectral analyses \citep{Montez+2010}.

\subsection{New X-ray-emitting Galactic PNe in the eROSITA eRASS1}

The inspection of the X-ray and optical images of possible eRASS1 counterparts of Galactic PNe has 
resulted in the discovery of X-ray emission from NGC\,2867 and IC\,1297.  These had not been targeted 
before by Chandra or XMM-Newton and are thus two new additions to the list of known X-ray-emitting PNe described below. 

\subsubsection{NGC\,2867}
NGC\,2867 (a.k.a.\ Caldwell\,90) is a prominent, high surface brightness PN with a faint, outer 
halo \citep[e.g.][]{1981MNRAS.197..647A} discovered by John Herschel in 1834. 
Its central star has been classified as a [WC3] with a terminal wind velocity $\simeq$~2000 km~s$^{-1}$ 
and the mass loss rate is estimated as $\simeq 3\times10^{-8}$ M$_\odot$~yr$^{-1}$ \citep{Keller+2014}. 

The location of the X-ray emission matches the optical nebula.   
The number of detected X-ray photons, $8.7\pm3.3$, is  
insufficient for a detailed spectral analysis, but we note that its eROSITA background-subtracted spectrum 
in Fig.~\ref{fig:spec_all} is relatively similar to that of NGC\,3242 while softer than that of NGC\,5315.  
Its median energy is $\sim0.55$ keV, which, according to its column density, is indicative of a 
plasma temperature $\simeq2\times10^6$~K (Fig.~\ref{fig:medeplot}).  
We thus propose it to be associated with a hot bubble.  

\subsubsection{IC\,1297}
IC\,1297 is a high surface brightness but little studied bipolar PN with a clear CSPN \citep{1986ApJ...311..930A}.  
Its morphology is actually reminiscent to that of NGC\,6543.  
Its CSPN is also classified as [WR] with spectral type [WO3] \citep{2003A&A...403..659A}.  

Just as for NGC\,2867, the location of the X-ray emission matches the optical nebula.  

The detected X-ray photons, $7.3\pm4.0$, are once again insufficient for detailed spectral analyses, 
but it is also noted that the background-subtracted spectrum of IC\,1297 in Fig.~\ref{fig:spec_all} 
is almost identical to that of NGC\,3242.  Furthermore, the median energy of these photons, $\sim0.44$ keV, 
in conjunction with the hydrogen column density towards IC\,1297, places it close to the locus of X-ray PNe 
with hot bubbles at a temperature $\simeq1\times10^6$~K (Fig.~\ref{fig:medeplot}).   
In view of these properties, we also propose it to be associated with a hot bubble.  

\subsection{Galactic PNe with spurious eROSITA eRASS1 counterparts}

The cross-correlation of the eRASS1 catalogue and HASH resulted in three additional potential 
matches, K\,1-27 (previously identified as an X-ray emitter), Wray\,16-385, and PHR\,J1107$-$5642.  
For various reasons, briefly outlined below, these are considered to be unreliable matches.  
Firstly, they exhibit none of the same optical nebula characteristics compared to the extant previously confirmed PNe that are 
also X-ray emitters such as having high surface brightness and prominent CSPN. 
Secondly, for both K\,1-27 and Wray\,16-385, there is a small but clear misalignment between the optical coordinate of the 
centre of the PN and the X-ray coordinate of its possible eRASS1 counterpart, which are 4.7-$\sigma$ and 6.0-$\sigma$ apart, 
respectively.  For the third case the detection of X-ray emission from PHR\,J1107$-$5642 is also highly dubious. 
Two factors are problematic in this case. These are the large angular size (188 arcsec in diameter) and 
optically faintness of the PN. These would be unprecedented for an X-ray emitting PN compared to all previously 
confirmed X-ray emitting PNe so we consider that the cross-correlation positional co-incidence is due to a
background source in close angular projection to the PN centroid. 

The spectral properties of the putative X-ray counterparts of K\,1-27 and Wray\,16-385 have median 
energies of 1.0 and 0.65 keV, respectively. The {\it eROSITA} median energy of K\,1-27 is higher 
than the typical value for PNe in Tables 1 and 2, making it a very unlikely X-ray value for a PN.  
For Wray\,16-385, its {\it eROSITA} median energy is also relatively high, although not completely 
incompatible with PNe or one with coronal emission from a companion of its central star.  
The X-ray spectra of these three sources in Fig.~\ref{fig:spec_all} clearly differ from those of 
confirmed X-ray-emitting PNe.  

The X-ray history of K\,1-27 is short, with limited previous literature available, but interesting 
to discuss.  \citet{1994A&A...286..543R} obtained a pointed ROSAT PSPC observation and claimed detection 
of X-ray emission from its central star.  \citet{Guerrero2000ROSAT} examined these same observations 
and cast doubt on the association between the X-ray emission and K\,1-27. This was because the 
X-ray emission spectral shape was not the typical of PNe while the positional match was not considered 
sufficiently robust. These doubts are now confirmed here with our {\it eROSITA} eRASS assessments leading 
to the conclusion that the X-ray emission is most likely a background source unrelated to K\,1-27.

\section{Discussion}
Below we provide a brief summary of the main findings from this {\it eROSITA} study of PNe.

\subsection{Frequency of Occurrence of X-ray-emitting PNe and Typical X-ray Flux}

The cross-correlation between all 1430 HASH `True' Galactic PNe that fall within the accessible 
{\it eROSITA}-DE eRASS1 source catalogue footprint ($179.94423568^\circ \leq l \leq 359.94423568^\circ$), 
resulted in a list of 10 PNe with possible eRASS1 candidate counterparts.  
Based on this work only seven can now be considered bona-fide eRASS1 PNe (Tabs.~\ref{tbl:pn_lst_ok} 
and \ref{tbl:pn_lst_new}).  This implies that of True HASH PNe only $\simeq$0.5\%, are detected 
in the eRASS1.  Their properties are listed in Table~\ref{tbl:pn_lst_review}.  The 11 true Galactic PNe 
previously confirmed as X-ray-emitters and falling in the available eRASS1 footprint but that are not 
detected by {\it eRosita} are also listed  in Table~\ref{tbl:pn_lst_review} together with their properties.  
Unsurprisingly all of them have X-ray fluxes below the nominal all-sky eRASS1 flux sensitivity limit 
of $5\times10^{-14}$ erg~cm$^{-2}$~s$^{-1}$. The eRASS1 PNe Abell\,30, NGC\,3242, and most 
likely NGC\,2867 also nominally fall below this base-level sensitivity limit, However, this varies across the 
sky, improving at high ecliptic latitudes which is the situation for most of these cases.  
It is certainly better for NGC\,2867, the faintest eRASS1 PN, which has an exposure time of 269~seconds, 
well above the average.  The comparison between X-ray PNe detected and undetected in eRASS1 in 
Table~\ref{tbl:pn_lst_review} can be used to set an average upper limit for the X-ray flux of the 1430 
True PNe in the Western, German region of eROSITA coverage. Comparing the detected Abell\,30 and 
undetected NGC\,4361, a threshold $\approx2\times10^{-14}$ erg~cm$^{-2}$~s$^{-1}$ can be determined.  
We can conclude that most (99.5\%) PNe in the Western, German footprint of eROSITA have X-ray fluxes 
below this threshold. Their low X-ray flux can be expected to be the result of a combination of distance, 
extinction, evolutionary status and inherent PN characteristics.  

\begin{table*}
\centering
\caption{X-ray-emitting Galactic PNe observed by eROSITA-DE in eRASS1.}
\label{tbl:pn_lst_review}
\tiny
\setlength{\tabcolsep}{3pt} 
\renewcommand{\arraystretch}{1.1}
\begin{tabular}{lccccccl}
\hline
PN Name &  d   & eRASS1 count rate  & Satellite & $f_{\rm X}$              & $T_{\rm X}$ & $L_{\rm X}$ & X-ray References \\ 
        & (pc) & (s$^{-1}$) &           & (erg~cm$^{-2}$~s$^{-1}$) & ($10^6$ K)  & (erg~s$^{-1}$) & \\

\hline
\multicolumn{8}{c}{Detected by eROSITA} \\
\hline
Abell\,30 & $2210\pm160$  & $0.071\pm0.036$ & XMM \& Chandra & $2.8\times10^{-14}$    & $0.79$ & $1.3\times10^{31}$ & 1 \\
IC\,1297  & $4400\pm1100$ & $0.091\pm0.038$ & eROSITA & $4.6\times10^{-14}$ & $\dots$  & $>1\times10^{32}$ & This paper \\
LoTr\,5   &  $455\pm10$   & $0.136\pm0.035$ & XMM, Chandra & $(9-23)\times10^{-14}$ & $7.5, 26-41$ & $(2.4-4.4)\times10^{30}$ & 2 \\
NGC\,2392 & $1830\pm90$   & $0.094\pm0.038$ & XMM, Chandra & $6\times10^{-14}$      & $2$ & $4.1\times10^{31}$ & 3, 4 \\
NGC\,2867 & $2900\pm500$  & $0.032\pm0.012$ & eROSITA & $1.6\times10^{-14}$ & $\dots$  & $>2\times10^{31}$ & This paper \\
NGC\,3242 & $1340\pm90$   & $0.083\pm0.036$ & XMM, Chandra & $4\times10^{-14}$      & $2.4$ & $4\times10^{31}$ & 5 \\
NGC\,5315 &  $960\pm190$  & $0.264\pm0.043$ & Chandra        & $1.0\times10^{-13}$    & $2.6$ & $4\times10^{31}$ & 6 \\ 
\hline
\multicolumn{8}{c}{Undetected by eROSITA} \\
\hline
DS\,1       &  $815\pm21$  & $0.011\pm0.005$ & Chandra & $9.5\times10^{-15}$ & $3.1, 14.5$ &  $1\times10^{30}$  & 2  \\
IC\,418     & $1360\pm50$  & $0.000\pm0.004$ & Chandra & $2.5\times10^{-15}$ & $3.0$       & $2.7\times10^{31}$ & 4 \\
Hb\,5       & 3200: & $0.040\pm0.009$ & XMM \& Chandra & $7.9\times10^{-15}$ & $2.4-3.7$   & $1.5\times10^{32}$ & 7 \\
HbDs\,1     &  $746\pm23$  & $0.014\pm0.006$ & Chandra & $5\times10^{-15}$   & $2.1$       & $4.7\times10^{29}$ & 8 \\
LO\,16      & $1820\pm150$ & $0.010\pm0.009$ & Chandra & $8.3\times10^{-16}$ & $15.9$      & $5.3\times10^{29}$ & 8 \\
NGC\,1360   &  $400\pm50$  & $0.018\pm0.004$ & Chandra & $1.4\times10^{-14}$ & $1.4$       & $2.8\times10^{29}$ & 8 \\
NGC\,2371-2 & $1720\pm140$ & $0.040\pm0.005$ & Chandra & $\dots$ & $\dots$ & $\dots$ & 9 \\
NGC\,4361   & $1040\pm50$  & $0.023\pm0.009$ & Chandra & $1.7\times10^{-14}$ & $0.67$      & $2.8\times10^{30}$ & 8 \\
NGC\,6153   & $1390\pm90$  & $0.015\pm0.005$ & Chandra & $\dots$ & $\dots$ & $\dots$ & 9 \\
NGC\,6337   & $1680\pm110$ & $0.019\pm0.009$ & Chandra & $7.0\times10^{-15}$ & $8.3-14.4$  & $1.8\times10^{31}$ & 8 \\
Sp\,1       & $1393\pm32$  & $0.007\pm0.004$ & Chandra & $1.3\times10^{-15}$ & $0.6-3.2$   & $1.2\times10^{30}$ & 8 \\
\hline
\multicolumn{8}{l}{
(1) \citet{Guerrero2012A30}; 
(2) \citet{Montez+2010};
(3) \citet{Guerrero2005XMMEskimo};
(4) \citet{Ruiz_2013NGC5315_eskimo};} \\
\multicolumn{8}{l}{
(5) \citet{Ruiz2011NGC3242}; 
(6) \citet{Kastner_2008NGC5315}; 
(7) \citet{Montez+2009}; 
(8) \citet{Montez+2015};} \\ 
\multicolumn{8}{l}{
(9) \citet{Freeman+2014}
}
\end{tabular}
\end{table*}

\subsection{The eRASS1 X-ray PNe}

\begin{figure}[htb!]
\centering
\includegraphics[width=0.5\textwidth]{ 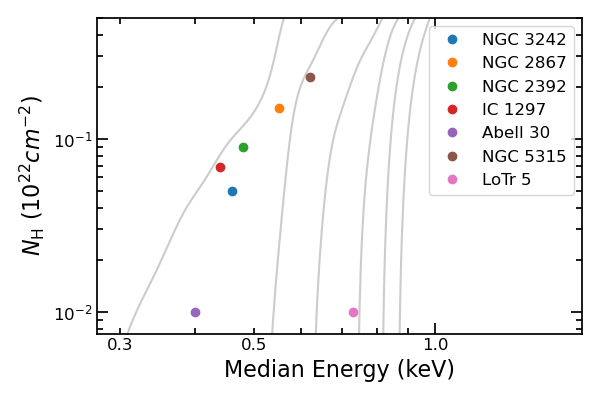}
\caption{
Median energy and column density plane for the eRASS1 X-ray PNe. Gray lines are for plasma 
temperatures of 1, 2, 3, 5, 8, and 10 MK increasing from the left to the right. 
}
\label{fig:medeplot}
\end{figure}

\begin{figure}[htb!]
\centering
\includegraphics[bb=40 40 550 550,width=0.5\textwidth]{ 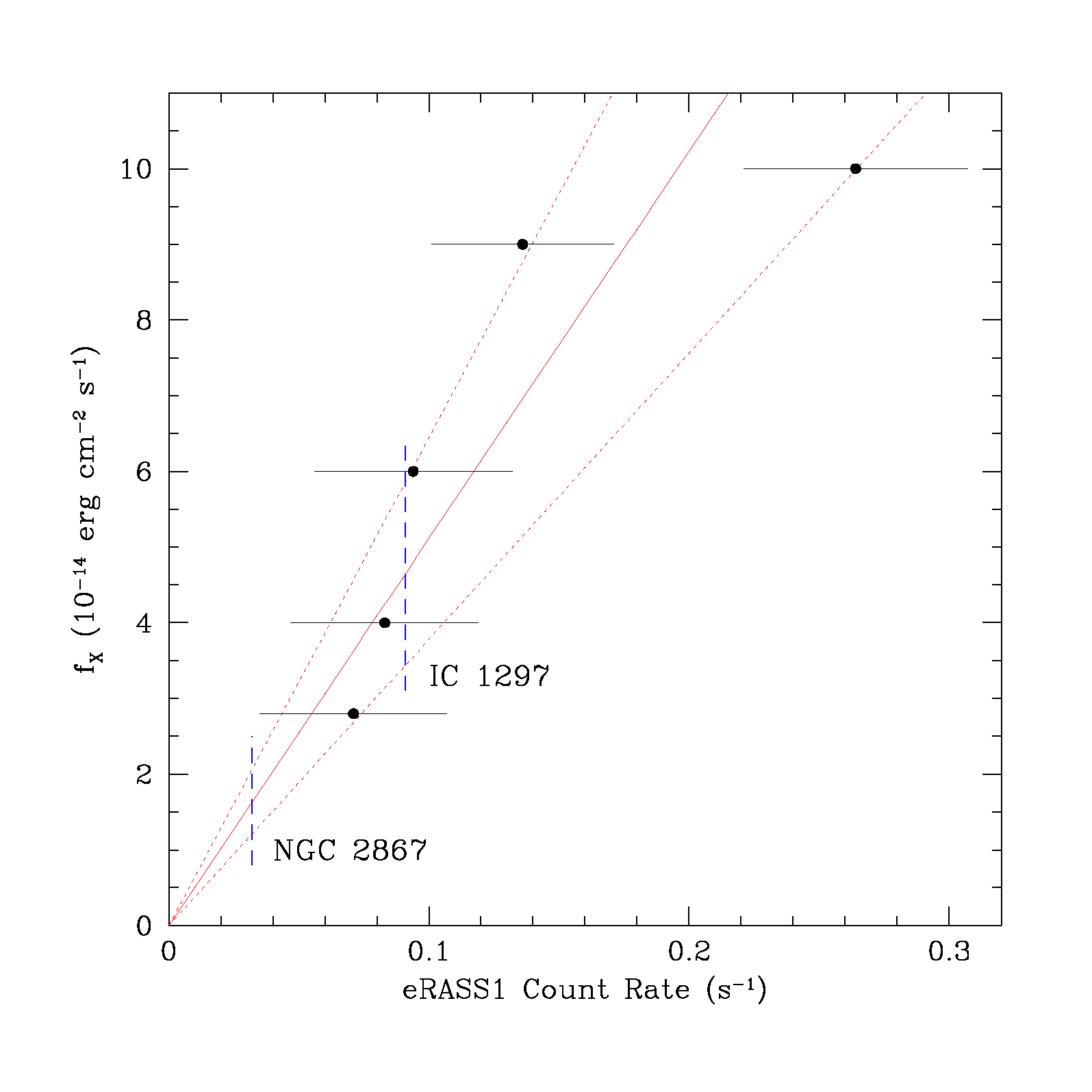}
\caption{
eROSITA count rate in the EML1 energy band and X-ray flux of the X-ray-emitting PNe. 
The red solid line corrresponds to the best-linear fit, whereas the red dotted lines consider 1-$\sigma$ dispersion.  
The count rates of IC\,1297 and NGC\,2867 are marked by vertical dashed blue lines. 
}
\label{fig:XPN}
\end{figure}

The eRASS1 catalogue includes the detection of five already known X-ray emitting PNe and two new detections.  
The nature of the five already known X-ray PNe detected by {\it eRosita} is varied: Abell\,30 is 
an extremely soft X-ray source, while NGC\,2392, NGC\,3242, and NGC\,5135 are hot bubbles of diffuse 
emission with varying plasma properties and extinction column densities. LoTr\,5 an apparently more evolved PN, 
is an X-ray point-source emitter arising from the interaction of the CSPN with a G-type companion.  
In general the spectral properties of X-ray PNe are different. 
This is also the case of the particular PNe detected by {\it eRosita}, as revealed in Figs.~\ref{fig:spec_all} 
and \ref{fig:medeplot}. Therefore a unique conversion factor from count rate to X-ray flux cannot be derived. 
To overcome this issue, the X-ray flux of these PNe, as derived from high quality Chandra or 
XMM-Newton spectra, are plotted in Figure~\ref{fig:XPN} against the observed eRASS1 EML1 band count rate.  
The red solid line shows the best linear-fit. The corresponding conversion factor from count rate to X-ray 
flux for PNe (red line in the plot) is found to be $(5.1\pm1.3)\times10^{-13}$ erg~cm$^{-1}$.  

No spectral fit is possible to the count-starved {\it eRosita} spectra of the new X-ray PNe IC\,1297 
and NGC\,2867. Their optical morphologies (right panels in Fig.~\ref{fig:pn_img_new}) and X-ray spectral 
properties (median photon energy in Tab.~\ref{tbl:pn_lst_review} and Fig.~\ref{fig:medeplot}
and X-ray spectra in Fig.~\ref{fig:spec_all}) are otherwise highly indicative of thermal plasma 
emission from hot bubbles. Using the conversion factor from eRASS1 EML1 count rate to X-ray flux, 
we derive X-ray fluxes of $(4.6\pm2.2)\times10^{-14}$ and $(1.6\pm0.7)\times10^{-14}$ erg~cm$^{-2}$~s$^{-1}$ 
for IC\,1297 and NGC\,2867, respectively. The intrinsic X-ray flux cannot be derived, but lower limits can 
be set from their observed X-ray fluxes and distances (see Tab.~\ref{tbl:pn_lst_review}).  

Both IC\,1297 and NGC\,2867 harbour [WR] CSPN of spectral type [WO3] and [WC2], respectively.  
This is relevant because it has been recognized in the past the existence of two different families 
of hot bubbles, those produced by H-rich CSPNe and those associated with [WR] CSPNe, with the more 
powerful stellar winds of the latter result in higher X-ray luminous PNe \citep{Kastner+2000,Freeman+2014,Toala+2019}. 
The X-ray luminosity of IC\,1297 is certainly high, typical of PNe with [WR] central stars.  
That of NGC\,2867, which is absorbed by a large column density, $\approx2\times10^{21}$~cm$^{-2}$, 
can also be expected to be high. These results further strengthen the X-ray expectations for PNe hosting [WR] central stars.  

\section{conclusions and expectations for new X-ray missions including the Einstein Probe}

The {\it eROSITA} eRASS1 catalogue has provided, for the first time, a survey of (half) the sky in X-rays at the sensitivity required to detect the soft and faint X-ray emission from hot bubbles of PNe and that of their CSPNe.  
Although 1430 True PNe from HASH with Galactic longitude $\gtrsim180^\circ$ are included in the footprint of eRASS1, 
only seven are detected, implying a low ($\approx0.5$\%) success rate, i.e., only one out of 200 PNe is found to exhibit X-ray 
emission at levels of a few times $10^{-14}$~erg~cm$^{-2}$~s$^{-1}$ that {\it eROSITA} is sensitive to.  
Future X-ray missions targeting PNe must have higher sensitivity to be capable of producing useful detections of a 
more statistically significant sample. Future {\it eROSITA} eRASS data releases and the availability of the Russian 
hemisphere {\it eRosita} data would be valuable to add further detections, 
but these questions are unfeasible for the time being. 

The new X-ray detections presented in this paper, namely IC\,1297 and NGC\,2867, both host [WR]-type CSPNe, 
confirming their superior levels of X-ray emission compared to PNe with H-rich CSPNe.  Pointed observations 
of these two PNe would be very useful to assess the properties of their X-ray-emitting gas and compare them 
with the nebular and stellar wind properties.  
We have begun an observing program with the Chinese Einstein Probe (EP) "lobster eye" X-ray telescope \citep{2015arXiv150607735Y, 2018SSPMA..48c9502Y, 2022hxga.book...86Y} to follow up some of the X-ray emitting PNe here but also others we have identified 
that have various common observational and physical characteristics that make them likely X-ray emitting PNe.
 The EP Wide-field X-ray Telescope features a 3600 sq. degree field of view, giving it the largest grasp among 
existing soft X-ray focusing telescopes. With an effective area of $600 cm^2$, decent angular resolution of $\leq30 ''$ 
(HPD), and a quick response time, the EP is among the most powerful X-ray telescopes. The strength of EP is in its 
rapid follow-up capability for observing transient X-ray sources, and it also works well for soft X-ray sources. 

The EP has similar base level performance in terms of spatial resolution/effective area to eROSITA, but unlike eROSITA, which operated in survey mode, the EP is an X-ray observatory that can make pointed and therefore deeper integrations up to tens of 
kilo-seconds if deemed necessary, reaching a sensitivity $\simeq10^{-14}$ erg~cm$^{-2}$~s$^{-1}$ in a 10 ks exposure \citep{Zhang2022}. 
At the same exposure level, Chandra ASIC and XMM-Newton have  flux limits of  $4\times 10^{-15}$ erg~cm$^{-2}$~s$^{-1}$ and $8.4\times 10^{-14}$ erg~cm$^{-2}$~s$^{-1}$ respectively. 

These new telescopes would enhance the detection of GRBs and supernovae, offering continuous power for exploring the universe.

\section{Acknowledgements}
We are grateful to the anonymous referee whose suggestions and feedback have improved the paper.
We would like to thank Dr. Detlef Schönberner for his careful reading of the manuscript and for catching several typographical errors.
We would like to thank the Einstein Probe Team for approving our proposal, which enabled the subsequent research presented in this work.
HY thanks QAP, HKU and the Laboratory for Space Research for the provision of a PhD scholarship.
MAG acknowledges financial support from grants CEX2021-001131-S funded by MCIN/AEI/10.13039/501100011033 
and PID2022-142925NB-I00 from the Spanish Ministerio de Ciencia, Innovaci\'on y Universidades (MCIU) 
co-funded with FEDER funds. QAP thanks the Hong Kong Research Grants Council for GRF research grants 17326116, 
17300417 and 17304520. We made use of NASA’s Astrophysics Data System;
The German eROSITA Consortium Data Release 1 (DR1);
the SIMBAD database operated at CDS, Strasbourg, France;
Astropy, a community-developed core Python package for Astronomy ~\citep{astropy};
Javalambre Photometric Local Universe Survey;
HASH, an online database at the Laboratory for Space Research at HKU federates available multi-wavelength imaging, spectroscopic;
and other data for all known Galactic PNe and is available at: http://www.hashpn.space.
This work is based on data from eROSITA, the soft X-ray instrument aboard SRG, a joint Russian-German science mission 
supported by the Russian Space Agency (Roskosmos), in the interests of the Russian Academy of Sciences represented 
by its Space Research Institute (IKI), and the Deutsches Zentrum für Luft- und Raumfahrt (DLR). The SRG spacecraft 
was built by Lavochkin Association (NPOL) and its subcontractors, and is operated by NPOL with support from the 
Max Planck Institute for Extraterrestrial Physics (MPE). The development and construction of the eROSITA X-ray 
instrument was led by MPE, with contributions from the Dr. Karl Remeis Observatory Bamberg \& ECAP (FAU Erlangen-Nuernberg), 
the University of Hamburg Observatory, the Leibniz Institute for Astrophysics Potsdam (AIP), and the Institute 
for Astronomy and Astrophysics of the University of Tübingen, with the support of DLR and the Max Planck Society. 
The Argelander Institute for Astronomy of the University of Bonn and the Ludwig Maximilians Universität Munich 
also participated in the science preparation for eROSITA.

\bibliography{sample7}{}
\bibliographystyle{aasjournalv7}

\end{document}